\documentclass[twocolumn,showpacs,preprintnumbers,amsmath,amssymb,nofootinbib]{revtex4-1}
\usepackage{epsfig}

\usepackage{comment}
\usepackage[table]{xcolor}
\usepackage{graphicx}
\usepackage{dcolumn}
\usepackage{bm}

\usepackage{hyperref}
\hypersetup{colorlinks=true,linkcolor=magenta,anchorcolor=green,citecolor=cyan,filecolor=black,menucolor=black,urlcolor=brown}
\usepackage{slashed}

\newcommand{\be}{\begin{equation}}
\newcommand{\ee}{\end{equation}}
\newcommand{\ba}{\begin{eqnarray}}
\newcommand{\ea}{\end{eqnarray}}

\begin{document}

\title{Detecting dark matter oscillations with gravitational waveforms}

\author{Philippe Brax}
\affiliation{Universit\'{e} Paris-Saclay, CNRS, CEA, Institut de physique th\'{e}orique, 91191, Gif-sur-Yvette, France}
\author{Clare Burrage}
\affiliation{School of Physics and Astronomy, University of Nottingham, Nottingham, NG7 2RD, United Kingdom}
\author{Jose A. R. Cembranos}
\affiliation{Departamento de F\'isica Te\'orica and IPARCOS, Facultad de Ciencias F\'isicas, \\
Universidad Complutense de Madrid, Ciudad Universitaria, 28040 Madrid, Spain}
\author{Patrick Valageas}
\affiliation{Universit\'{e} Paris-Saclay, CNRS, CEA, Institut de physique th\'{e}orique, 91191, Gif-sur-Yvette, France}

\begin{abstract}

We consider the phase shift in the gravitational wave signal induced by  fast oscillations of  scalar dark matter surrounding   binary systems, which could be probed
by the future experiments LISA and DECIGO. This effect depends on the local matter density and the mass of the dark matter particle. We compare it to the phase shift due to a standard dynamical friction term,
which should generically be present. We find that the effect associated with the oscillations only
dominates over the dynamical friction for dark matter masses below $10^{-21}$ eV, with
masses below $10^{-23}$ eV  implying cloud sizes that are too large to be realistic. Moreover,
for masses of the order of $10^{-21}$ eV, LISA and DECIGO would only detect this effect for
dark matter densities greater than that in the solar system by a factor $10^5$ or $10^4$ respectively. We conclude that this signal can be ignored for most dark matter scenarios unless very dense clouds of very light dark matter are created early in the Universe at a redshift $z\sim 10^4$.

\end{abstract}

\maketitle

\section{Introduction}
\label{sec:introduction}

The exploration of dark matter (DM) halos as scalar field solitons of extended sizes has emerged as a promising avenue to test the fundamental nature of DM. Recent studies have investigated this  possibility as an alternative framework addressing some of the small-scale observational challenges of  the conventional Cold DM scenario \cite{Hui:2016ltb, Bar:2018acw}. Indeed Scalar DM models, distinguished by a background oscillating field whose pulsation is determined by the mass of the scalar particle, introduce a novel perspective on the dynamics of DM \cite{Cembranos:2015oya}. Although these oscillations manifest themselves at rapid rates on cosmological and astrophysical scales, a common analytical strategy involves integrating them out, focusing on the slow temporal and spatial variations in the amplitude of the scalar field for a more manageable analysis \cite{Turner:1983he, Johnson:2008se, Brax:2019fzb}.

The framework used in this work includes the popular fuzzy DM (FDM) models as an alternative to the traditional cold DM paradigm \cite{Hu2000}. FDM, characterized by ultralight bosonic DM constituents within the mass range $10^{-22} - 10^{-20}$ eV, has garnered attention due to its unique description based on a single coherent wavefunction \cite{Schive2014, Mocz2017}. Numerical simulations have demonstrated that FDM, on large scales, reproduces the cosmic web structure of CDM while addressing challenges faced by CDM on galactic scales \cite{Schive2014, Mocz2017, May2021StructureDynamics}. Favourable aspects of FDM, such as resolving the cusp-core problem \cite{Maccio1}, have led to an increased interest in its potential to remedy purported issues with the CDM model \cite{2015PNAS..11212249W, 2017ARA&A..55..343B, 2017Galax...5...17D}. This has sparked extensive research into the dynamics of FDM solitons, also known as cores, within halos \cite{Mocz2019, Mocz2020, Banerjee2020, Mina2022}. The cores in FDM halos exhibit intriguing dynamical behaviors, including random walks within the gravitational potential  and stochastic features \cite{Li2021, Hui2021-2, Woo1, Schive2014, Mocz2017, Marsh2019, Veltmaat2018, Glennon2020-2}, as well as homogeneous radial expansions and contractions of the soliton core, sending out density waves into the surrounding halo \cite{Woo1}.

Historically, most investigations of FDM have considered non-interacting bosons, described by a Schrödinger equation coupled to the Poisson equation in the Schrödinger-Poisson system of coupled equations (SPE) \cite{Chavanis2011, Chavanis2011_2, Rindler-Daller:2011afd, Li2014, Li2017, Desjacques:2017fmf, Chavanis:2020rdo, Hartman2022, Hartman2022-2, Mocz_2023, Chakrabarti2022, Dave2023,Dave:2023egr}. However, recent efforts have explored the  dynamical effects  introduced by interactions, which add nonlinear contributions to the Schrödinger equation \cite{Chavanis2011, Chavanis2011_2, Rindler-Daller:2011afd, Li2014, Li2017, Desjacques:2017fmf, Chavanis:2020rdo, Hartman2022, Hartman2022-2, Mocz_2023, Chakrabarti2022, Dave2023}. Such interactions lead to   a more complex and dynamically rich scenario. After integrating out the fast oscillations of the scalar field,  the dynamics of DM can be understood as the one of a fluid, where  equilibria arise from the balance  between gravity, self-interactions, and the `quantum-pressure' resulting from the spatial gradients of the scalar. These solitons could play  a significant role in shaping the large-scale structure of the universe. The Fuzzy DM solitons are associated to  the  balance between quantum-pressure and gravity \cite{Chavanis:2011uv}. Other solitons  where repulsive self-interactions equilibrate with gravity provide alternative examples \cite{Brax:2019fzb}.

In this article, we consider  scalar field solitons and their influence on  astrophysical phenomena, in particular  the propagation of gravitational waves (GW).  Small solitons, whose sizes could be significantly smaller than galactic halos and could be formed during different cosmic eras, e.g the matter or radiation eras \cite{Brax:2020oye}, could modify the phase of GW produced by binary systems. These solitons, exhibiting densities much larger than the average DM density in galactic halos, may constitute a significant portion of the total DM in the universe. For instance, if  formed around  the time of matter-radiation equality, they could reach densities as high as one million times the local DM density in the solar system.  We specifically focus on binary systems that could potentially belong to these dense clumps and the effects on the GWs they produce that could be  probed by  future experiments like LISA 
\cite{Danzman-1996,LISA:2017pwj,LISA:2022yao} 
and DECIGO \cite{Seto-2001,Nakamura-2016,Kawamura:2020pcg}. 
We note that  the capture of binary systems by these dense clumps remains a subject for future study.

Environmental effects, associated with baryons or dark matter, generically affect the gravitational
waves signal emitted by binary systems via their impact on the orbital features. 
Typically, they are the consequences of three possible effects, 
the conservative gravitational pull of the enclosed dark matter mass within the binary orbit,
the dynamical friction (i.e., the drag force on the binary components due to the gravitational
exchange of momentum with the environment), or the accretion of matter.
Recent studies of these effects can be found for instance in 
\cite{Vicente:2022ivh,Cardoso:2019rou,Bamber:2022aa,Traykova:2023aa,Boudon:2023vzl,Aurrekoetxea:2023aa}.
In this paper, following \cite{Khmelnitsky:2013lxt}, we focus on a different effect that is
specific to scalar field scenarios (as opposed to classical particles or fluids) and associated
with the fast oscillations of the scalar field $\phi(\vec x,t)$, which can be written as
\be
\phi(\vec x, t) = A(\vec x, t) \; \cos[ m_\phi t + \alpha(\vec x,t) ] .
\label{eq:phi-cos}
\ee
where $m_\phi$ is the dark matter particle mass.
The amplitude $A$ and the phase $\alpha$ vary on astrophysical or cosmological timescales. 
Similarly the dark matter density and the gravitational field, when averaged over the fast oscillations of the scalar field at frequency
$m_\phi$, vary also on such large time scales. As recalled above, the dynamics of the dark matter density field, such as the formation
of solitons, are usually studied using the equations of motion obtained after this averaging procedure,
written in terms of the amplitude $A$ and the phase $\alpha$ or the complex field $\psi = A e^{i\alpha}$.
However, as pointed out by \cite{Khmelnitsky:2013lxt}, the underlying fast oscillations
(\ref{eq:phi-cos}) lead to a subleading oscillating component of the gravitational potential
$\Psi_N$, as in Eq.(\ref{eq:Psi-N-cos}) below.
This in turns gives rise to a specific time-dependent shift of the gravitational waveform,
which is not due to a change of the orbital dynamics but to the propagation of the gravitational
wave in the surrounding oscillating gravitational potential.

In this paper, we compute this specific phase shift and we compare its magnitude with a generic dynamical friction \cite{Chandrasekhar:1943ys, Barausse:2014tra, Boudon:2023vzl}. Interestingly, we find that the oscillating DM effects can only be probed for a specific range of scalar masses, dependent on the GW frequency and the total mass of the binary system. Practically, our results suggest that only scalar masses lower than $10^{-21}$ eV could be tested when the local matter density exceeds one million times the estimated density for DM in the Milky Way. The formation of very dense clumps around the matter-radiation equality epoch would then lead to potentially observable effects.
Our analysis applies to scenarios of the form (\ref{eq:phi-cos}), which include both FDM models
and models with non-negligeable self-interactions.

In Section \ref{sec:GWshift}, we compute the shift in the frequency and phase of the gravitational waves due to the oscillating dark matter background, and compare this to the size of similar effects arising from dynamical friction. In Section \ref{sec:detec}, we use a Fisher matrix analysis to determine which local dark matter overdensities can be probed with near future experiments, focusing in particular on LISA and DECIGO.  We conclude in Section \ref{sec:conc}.

\section{Impact of the time-dependent dark matter potential on GW}
\label{sec:GWshift}
\subsection{Frequency shift}

In the near  future the LISA experiment will detect and analyse the GWs due to white dwarf binaries in the Milky Way. It is expected that over 10 years of observation, some $10^4$ White Dwarf Binaries (WDB) will be observed at frequencies $f_0\gtrsim 5$ mHz \cite{Lamberts:2019nyk,Seto:2022iuf}. These systems will allow us to test the nature of their DM environment \cite{Khmelnitsky:2013lxt}.
In a fashion similar to the Sachs-Wolfe effect for the Cosmic Microwave Background (CMB),
the fluctuations of the gravitational potential along the line of sight lead to a
drift of the frequency, $f$, of the emitted GWs,
\be
\frac{\Delta f}{f} =  \Psi_N(\vec x_e,t_e) - \Psi_N(\vec x, t)  ,
\label{eq:Sachs}
\ee
where $\{\vec{x}_e,t_e\}$ and $\{\vec{x},t\}$ indicate the position and time of the
emission and reception of the GW.
This description is valid as long as the GWs can be described as rays in an optical approximation where their frequency must be larger than the inverse of
the typical variation scale of the surrounding medium.
The integrated Sachs-Wolfe effect is also neglected. This follows from the fact that the spatial variation of the gravitational potential occurs on scales much larger than the wavelength.

If the galactic halo is composed of clumps of DM whose particle mass is $m_\phi$, the local matter density will include a subleading component that oscillates with a frequency $\omega=2 m_\phi$ locally inside each clump, associated with the underlying oscillation (\ref{eq:phi-cos}) of the field.
Through the Einstein equations we find the  local Newtonian potential to be \cite{Khmelnitsky:2013lxt}
\be
\Psi_N(\vec x,t) = \Psi_0(\vec x) + \Psi_{\rm osc}(\vec x) \cos[\omega t + 2 \alpha(\vec x) ] ,
\label{eq:Psi-N-cos}
\ee
with
\be
\omega = 2 m_\phi .
\ee
The leading component in Eq.~(\ref{eq:Psi-N-cos}), $\Psi_0$, which evolves on astrophysical timescales, is given by
the usual Poisson equation,
\be
\nabla^2 \Psi_0 = 4 \pi {\cal G} \rho ,
\label{eq:Psi-0-rho}
\ee
where $\rho$ is the DM density averaged over the fast oscillations at frequency
$\omega$, whereas the subleading oscillating component $\Psi_{\rm osc}$ is given by
\be
\Psi_{\rm osc} = \pi \frac{{\cal G} \rho}{m_\phi^2} .
\label{eq:Psi-osc-rho}
\ee
The de Broglie wavelength $\lambda_{\rm dB}$ of the DM particles is
$\lambda_{\rm dB} = 2\pi/(m_\phi v)$, with $v$ the typical virial velocity of the DM
cloud.
The effective quantum pressure smoothes out inhomogeneities on scales smaller than
$\lambda_{\rm dB}$, therefore typical wavenumbers $k$ of the DM density field
verify $k < 2\pi/\lambda_{\rm dB}$ ($k$ can be much smaller if there are repulsive
self-interactions that contribute to an additional pressure, or more generally as in
CDM scenarios when the size of the cloud is related to its formation process rather than
to $m_\phi$).
Then, comparing equations (\ref{eq:Psi-0-rho}) and (\ref{eq:Psi-osc-rho}) we have
\be
k < \frac{2\pi}{\lambda_{\rm dB}} : \;\;\; k < m_\phi v , \;\;\;
\frac{\Psi_{\rm osc}}{\Psi_0} \sim \frac{k^2}{m_\phi^2} < v^2 \ll 1 ,
\label{eq:Psi-osc-Psi0}
\ee
for nonrelativistic DM clouds.

As pointed out in Ref.~\cite{Khmelnitsky:2013lxt} in the context of Pulsar Timing Arrays (PTAs), the oscillating component $\Psi_{\rm osc}$
will lead, through Eq.~(\ref{eq:Sachs}), to an oscillating frequency drift of the GW,
which could be detected, whereas the constant term $\Psi_0$ is degenerate with binary parameters.
We shall find below that a detection requires a DM density that is much larger
than the solar neighborhood estimate. Therefore, we can assume the gravitational potential at emission
to dominate in Eq.~(\ref{eq:Sachs}) and we write the observed frequency of the GW signal as
\be
f= \bar f + \Delta f = \bar f (1 + \Psi) ,
\label{eq:f-fbar-Psi}
\ee
where $\bar f$ is the unperturbed frequency, that is, for a binary system in vacuum, and
$\Delta f$ is the frequency shift due to the binary DM environment, with
\be
\Psi = \Psi_0 + \Psi_{\rm osc} \cos(\omega t + 2 \alpha) ,
\label{eq:Psi-Psi0-Psi-osc}
\ee
the potential at emission.
The optical approximation (\ref{eq:Sachs}) is valid for
\be
f \gtrsim \omega , \;\;\; \mbox{whence} \;\;
m_\phi < \left (\frac{f_{\rm min}}{1 \, {\rm Hz}} \right )\; 3 \times 10^{-16} \, {\rm eV} ,
\label{eq:f-omega-opt}
\ee
where $f_{\rm min}$ is the minimum frequency of the GW interferometer.
Compared with the contributions from Eq.~(\ref{eq:Sachs}), the integrated Sachs-Wolfe effect is
suppressed by a factor $k/\omega < v \ll 1$ and can be neglected for nonrelativistic clouds.

Throughout this paper, we work at linear order in the DM density and gravitational
potential. Our analysis is not restricted to the clouds associated with solitons
in Fuzzy DM scenarios (i.e., stable equilibria governed by the balance between gravity
and quantum pressure). It also applies to more general cases, such as solitons governed by the
balance between gravity and the effective pressure due to repulsive self-interactions,
or virialized halos supported by their velocity dispersion (as for CDM).

\subsection{Gravitational wave phase shift}
\label{sec:GW-phase-shift}

The GW signal from the binary systems we consider takes the form $h(t) = A(t) \cos[\Phi(t)]$, where the phase $\Phi(t)$ and the time $t$
are related to the frequency $f$ and the frequency drift $\dot f$ by
\be
\Phi = 2\pi \int df \frac{f}{\dot f} ,  \;\;\; t = \int df \frac{1}{\dot f} .
\ee
At leading order, the amplitude grows as $A(t) \propto f^{2/3}$ and the frequency drift due to the emission
of GW by the binary system reads
\be
\dot f = \frac{96 \pi^{8/3}}{5 c^5} ( {\cal G} {\cal M})^{5/3} f^{11/3} ,
\label{eq:dot-f-GW}
\ee
where ${\cal M}$ is the chirp mass of the two compact objects of mass $m_1$ and $m_2$, and
\be
M= m_1+m_2 , \;\;\; \nu=m_1 m_2/M^2 , \;\;\; {\cal M} = \nu^{3/5} M ,
\ee
where $\nu$ is the symmetric mass ratio \cite{poisson_will_2014,Barausse:2014tra}.

Going to Fourier space,
$\tilde h(f) = \int dt e^{i2\pi f t} h(t)$, one obtains in the stationary phase approximation
$\tilde h(f) = A(f) e^{i\psi(f)}$ with
\be
A(f) \propto f^{-7/6} , \;\;\;
\psi(f) =  2\pi f t_\star - \Phi(t_\star) - \pi/4 ,
\ee
where the saddle-point $t_\star$ is determined by $f(t_\star) = f$.

At zeroth order in the DM environment, we have
$\bar f(\bar t_\star) = f$ and
\be
t_c - \bar t_\star = \int_f^{\infty} d f \frac{1}{\dot f}
= \frac{5}{256 \pi} \left( \frac{\pi {\cal G} {\cal M}}{c^3} \right)^{-5/3} f^{-8/3} ,
\label{eq:bar-t-f}
\ee
with the phase  given by
\be
\Phi_c - \bar\Phi_\star = 2\pi \int_f^{\infty} d f \frac{f}{\dot f}
= \frac{1}{16} \left( \frac{\pi {\cal G} {\cal M} f}{c^3} \right)^{-5/3} ,
\ee
where $t_c$ and $\Phi_c$ are the time and the phase at that  coalescence time.
This gives the standard result for the phase $\bar\psi(f)$ of the Fourier-space waveform:
\be
\bar\psi(f) = 2\pi f t_c - \Phi_c - \frac{\pi}{4} + \psi_{\rm GW}(f)
\label{eq:bar-psif}
\ee
with
\ba
\psi_{\rm GW}(f) & = & \frac{3}{128} \left( \frac{\pi {\cal G} {\cal M} f}{c^3} \right)^{-5/3}
\biggl[ 1 + \left( \frac{3715}{756} + \frac{55 \nu}{9} \right) \nonumber \\
&& \times \left( \frac{\pi {\cal G} M f}{c^3} \right)^{2/3} \biggl] .
\label{eq:psi-GW}
\ea
Here we have included the first post-Newtonian correction (1-PN order).
This gives two terms, which behave as $f^{-5/3}$ and $f^{-1}$, that allow us to constrain both binary
masses $m_1$ and $m_2$ from the observations \cite{Boudon:2023vzl}. We do not consider higher order post-Newtonian
contributions in this paper, which can be used to constrain the spins of the compact objects.

Because of the DM perturbation Eq.~(\ref{eq:f-fbar-Psi}), the saddle-point time $t_\star$
associated with a frequency $f$ is shifted at first order by
\be
t_\star = \bar t_\star + \Delta t_\star , \;\;\; \mbox{with} \;\;\;
\Delta t_\star = - \frac{\bar f(\bar t_\star)}{\bar f'(\bar t_\star)} \Psi(\bar t_\star) ,
\ee
while the phase $\Phi_\star= \bar \Phi_\star + \Delta \Phi_\star$ is shifted by
\be
\Delta \Phi_\star = 2\pi f \Delta t_\star - 2\pi \int_{\bar t_\star}^{t_c} dt \bar f \Psi .
\ee
This gives a shift of the phase $\Delta\psi(f)$ of the Fourier-space waveform
\be
\Delta\psi(f) = 2\pi \int_{\bar t_\star}^{t_c} dt \bar f \Psi .
\ee
Using Eq.~(\ref{eq:bar-t-f}), we can write this integral as
\be
\Delta\psi = 2\pi \left( \frac{5}{256 \pi} \right)^{3/8} \left( \frac{\pi {\cal G} {\cal M}}{c^3} \right)^{-5/8}
\int_{\bar t_\star}^{t_c} dt (t_c-t)^{-3/8} \Psi(t) .
\ee
The constant term $\Psi_0$ of the gravitational potential in Eq.~(\ref{eq:Psi-Psi0-Psi-osc}) gives the contribution
\be
\Delta\psi_0(f) = \frac{\Psi_0}{16} \left( \frac{\pi {\cal G} {\cal M} f}{c^3} \right)^{-5/3} .
\label{eq:psif-0}
\ee
We can see that this term, which scales as $f^{-5/3}$, is fully degenerate with the leading  GW phase in Eq.~(\ref{eq:psi-GW}). Moreover, for nonrelativistic DM clouds
$\Psi_0 \ll 1$. Therefore, we would need to know the distribution of white dwarf masses
(or more generally binary masses) with a very high accuracy to distinguish the effect
of the contribution (\ref{eq:psif-0}).
Thus $\Psi_0$ cannot be discriminated from a small shift of the binary masses
$m_1$ and $m_2$ and we do not consider it any further.

The time-dependent term of the gravitational potential in Eq.~(\ref{eq:Psi-Psi0-Psi-osc}) gives the contribution
\ba
\Delta\psi_{\rm osc}(f) & = & \Psi_{\rm osc} 2\pi \left( \frac{5}{256 \pi} \right)^{3/8}
\left( \frac{\pi {\cal G} {\cal M} \omega}{c^3} \right)^{-5/8} \nonumber \\
&& \times {\rm Re}\left[ e^{i (5\pi/16+\theta-\omega t_c)} \, \gamma(5/8,-i y) \right] ,
\label{eq:Psi-osc-gamma}
\ea
where $\gamma(a,z)$ is the incomplete gamma function and
\be
y = \omega (t_c-\bar t_\star) = \frac{m_\phi}{m_\star} , \;\;\;
m_\star = f \frac{128 \pi}{5} \left( \frac{\pi {\cal G} {\cal M} f}{c^3} \right)^{5/3} .
\label{eq:M-star-def}
\ee

For low scalar mass, $m_\phi \ll m_\star$, we obtain
\be
m_\phi \ll m_\star : \;\; \Delta\psi_{\rm osc}(f) = \frac{\Psi_{\rm osc}}{16}
\left( \frac{\pi {\cal G} {\cal M} f}{c^3} \right)^{-5/3} \cos(\omega t_c-\theta) .
\ee
Because of the bounds in Eq.~(\ref{eq:Psi-osc-Psi0}), this phase shift is even smaller than
for the constant potential contribution of Eq.~(\ref{eq:psif-0}) and it is again degenerate with the
GW phase in Eq.~(\ref{eq:psi-GW}).
For high scalar masses, $m_\phi \gg m_\star$, we obtain
\ba
&& m_\phi \gg m_\star : \;\; \Delta\psi_{\rm osc}(f) = \Psi_{\rm osc} \Gamma(5/8) 2\pi
\left( \frac{5}{256 \pi} \right)^{3/8} \nonumber \\
&& \times \left( \frac{\pi {\cal G} {\cal M} \omega}{c^3} \right)^{-5/8} \cos(\omega t_c-\theta-5\pi/16) ,
\ea
which is degenerate with the constant factor $\Phi_c$ in Eq.~(\ref{eq:bar-psif}).
Therefore, the DM phase shift is degenerate in both low and high scalar mass limits.
This means that the contribution of the  DM environment binary gravitational wave forms can only potentially be distinguished for scalar masses
of the order of $m_\star$, which can span a few orders of magnitude depending on the
frequency range of the GW interferometer. { Notice that this typical mass $m_\star$ is much smaller than the signal's frequency as
long as the GWs do not probe the Schwarzschild radius of the system. This must of course be satisfied  for our semi-classical description
of the propagation of the GWs to hold. }

The factor (\ref{eq:Psi-osc-gamma}) depends on the chirp mass ${\cal M}$, which at the Newtonian level is degenerate with $\Psi_0$ as seen in (\ref{eq:psif-0}). However, for nonrelativistic DM clouds
$\Psi_0 \ll 1$ and this shift can only lead to a small bias in the measurement of ${\cal M}$,
as seen by the comparison with (\ref{eq:psi-GW}).
Therefore, we can neglect the impact of the shift (\ref{eq:psif-0}) and the parameter
$\Psi_0$ in the Fisher matrix analysis described in Sec.~\ref{sec:detec} below.
For $|\Psi_0| < 0.1$ the detection thresholds that we obtain for $\Psi_{\rm osc}$ and the dark matter
density $\rho$ would be biased by less than $10\%$.

\subsection{Comparison with dynamical friction}

If a binary system is embedded within a DM halo, its GWs signal will be
affected by other, more usual, effects, in addition to the phase shift in Eq.~(\ref{eq:Psi-osc-gamma})
associated with the specific oscillatory behavior of the Newtonian potential in Eq.~(\ref{eq:Psi-N-cos}).
These include the impact of the DM halo on the orbital radius of the binary, due to
gravitational force from the enclosed DM mass, the matter accretion onto the compact
objects, and the dynamical friction.
The rate of matter accretion can depend on the details of the DM model but the dynamical
friction often takes the form of the usual Chandrasekhar result  \cite{Chandrasekhar:1943ys}
\be
m_i \dot{\vec v}_i = - \frac{4\pi {\cal G}^2 m_i^2 \rho}{v_i^3} \Lambda \vec{v}_i ,
\label{eq:dyn-fric}
\ee
where $\Lambda$ is the Coulomb logarithm and the index $i=\{1,2\}$ labels the two components of the
binary system.
The expression in Eq.~(\ref{eq:dyn-fric}) derived for collisionless media, such as CDM, also applies to
Fuzzy DM or scenarios with non-negligible self-interactions in the supersonic regime,
although $\Lambda$ depends on the model.
Therefore, it is interesting to compare the the phase shift we derived in Eq.~(\ref{eq:Psi-osc-gamma}) with the
generic effect of the dynamical friction, Eq.~(\ref{eq:dyn-fric}), which is expected to be also present
in most cases.
To keep the analysis general and simple, we approximate $\Lambda$ as a constant, and
in our numerical computations we will take $\Lambda=10$.
As described for instance in Ref.~\cite{Boudon:2023vzl}, the drag force, Eq.~(\ref{eq:dyn-fric}), leads to
a slow decay of the orbital radius $a$, in addition to the shrinking due to the emission of
GWs, which reads
\be
\dot a_{\rm df} = - a \left(\frac{a}{{\cal G} M} \right)^{3/2} 8 \pi {\cal G}^2 \rho \Lambda
\frac{m_1^3+m_2^3}{\mu^2} ,
\ee
where $\mu=m_1 m_2/M$ is the reduced mass.
This in turn gives rise to an additional drift of the GWs frequency;
\be
\dot f_{\rm df} = 12 {\cal G}\rho \frac{\Lambda (m_1^3+m_2^3)}{\nu^{1/5} {\cal M}^3} ,
\label{eq:dot-f-df}
\ee
and to a phase shift;
\be
\Delta\psi_{\rm df} = - \frac{75}{38912} \frac{\pi {\cal G}^3 {\cal M} \rho}{c^6}
\left( \frac{\pi {\cal G} {\cal M} f}{c^3} \right)^{-16/3}
\frac{\Lambda (m_1^3+m_2^3)}{\nu^{1/5} {\cal M}^3} .
\label{eq:psi-df}
\ee
Here, as in \cite{Boudon:2023vzl}, we consider the effects due to DM as a linear perturbation to the GW emission   and assumed that
the contribution of Eq.~(\ref{eq:dot-f-df}) to the frequency drift is small as compared with the
contribution of Eq.~(\ref{eq:dot-f-GW}) due to the emission of GWs.

\section{Detection threshold}
\label{sec:detec}

\subsection{Fisher matrix analysis}

We use a Fisher matrix analysis to investigate which DM densities can be probed
by GW waveforms, through the impact of the oscillating Newtonian potential in Eq.~(\ref{eq:Psi-Psi0-Psi-osc})
on the phase of Eq.~(\ref{eq:Psi-osc-gamma}).
As usual \cite{Poisson:1995ef,Vallisneri:2007ev}, the Fisher matrix reads
\be
\Gamma_{ij} = 4 \, {\rm Re} \int_{f_{\min}}^{f_{\max}} \frac{df}{S_n(f)} \,
\left( \frac{\partial\tilde h}{\partial\theta_i} \right)^\star
\left( \frac{\partial\tilde h}{\partial\theta_j}\right) ,
\ee
where $S_n(f)$ is the noise spectral density of the GW interferometer and
$\{\theta_i\}$ is the set of parameters that we wish to measure.
In this paper we consider $\{\theta_i\} = \{t_c, \Phi_c, \ln(m_1), \ln(m_2), \Psi_{\rm osc} \}$,
as we discard the spins of the compact objects.
The amplitude ${\cal A}_0$ would be an additional parameter, however, the Fisher matrix
is block-diagonal as $\Gamma_{{\cal A}_0,\theta_i}=0$ and the amplitude ${\cal A}_0$
is completely decorrelated from the other parameters $\{\theta_i\}$ \cite{Poisson:1995ef}.
Therefore, we do not consider the amplitude any further.
 From the Fisher matrix $\Gamma_{ij}$ we obtain the covariance matrix
$\Sigma_{ij} = \left(\Gamma^{-1}\right)_{ij}$, which gives the standard deviation on the
various parameters as
$\sigma_i = \langle (\Delta\theta_i)^2 \rangle^{1/2} = \sqrt{\Sigma_{ii}}$.
We obtain in this fashion the 1-sigma error bar on the amplitude of the DM
oscillating potential $\Psi_{\rm osc}$, or equivalently on the DM density $\rho$
through Eq.~(\ref{eq:Psi-osc-rho}).
We perform the analysis for a fiducial $\rho=0$, i.e. assuming the binary is in vacuum.
Then, we identify $\sigma_\rho$ as the detection threshold on the DM density
$\rho$.

The signal-to-noise ratio is given by
\be
({\rm SNR})^2 = 4 \int_{f_{\min}}^{f_{\max}} \frac{df}{S_n(f)} \, | \tilde h(f) |^2 .
\ee
Writing the GWform as $\tilde{h}(f) = {\cal A}_0 f^{-7/6} e^{i\psi(f)}$ at leading
order, we obtain the standard expression
\be
\Gamma_{ij} = \frac{({\rm SNR})^2}{\int_{f_{\min}}^{f_{\max}} \frac{df}{S_n(f)} f^{-7/3}}
\int_{f_{\min}}^{f_{\max}} \frac{df}{S_n(f)} f^{-7/3} \frac{\partial\psi}{\partial\theta_i}
\frac{\partial\psi}{\partial\theta_j} .
\label{eq:Fisher-def}
\ee
The derivatives are computed from Eqs.(\ref{eq:bar-psif}), (\ref{eq:psi-GW}) and
(\ref{eq:Psi-osc-gamma}), which we simplify as
\ba
\Delta\psi_{\rm osc}(f) & \sim & \Psi_{\rm osc} 2\pi \left( \frac{5}{256 \pi} \right)^{3/8}
\left( \frac{\pi {\cal G} {\cal M} 2 m_\phi}{c^3} \right)^{-5/8} \nonumber \\
&& \times \left|  \gamma \left( \frac{5}{8} , -i \frac{m_\phi}{m_\star(f)} \right) \right| ,
\label{eq:Psi-osc-gamma-abs}
\ea
where we have discarded the random phase factor and directly compute the modulus of the
term in the real part.
This provides the order of magnitude of the phase shift associated with the DM
environment, through the impact of the oscillating gravitational potential.
As we take $\Psi_{\rm osc}=0$ as the fiducial case, the derivatives in Eq.~(\ref{eq:Fisher-def}) with
respect to $\{t_c, \Phi_c, \ln(m_1), \ln(m_2)\}$ arise from the zeroth-order terms, Eqns.~(\ref{eq:bar-psif}) and (\ref{eq:psi-GW}), whereas the derivative with respect to
$\Psi_{\rm osc}$ is given by (\ref{eq:Psi-osc-gamma-abs}).

To compare with the impact  of dynamical friction, Eq.~(\ref{eq:psi-df}), we also perform a separate
Fisher analysis where we only include the phase shift in Eq.~(\ref{eq:psi-df}) for the impact of DM.
Then we consider the parameters $\{\theta_i\} = \{t_c, \Phi_c, \ln(m_1), \ln(m_2), \rho \}$
and we directly obtain the detection threshold $\sigma_\rho$ on the DM density
$\rho$, which does not depend on the particle mass $m_\phi$.

\subsection{LISA}

\begin{table}[hbtp]
\centering
\begin{tabular}{|l||*{5}{c|}}
 \hline
 \rule{0pt}{10pt} & $m_1$ ($\rm M_{\odot}$) & $m_2$ ($\rm M_{\odot}$) & SNR & $d_L$ (Mpc) & detections \\ [0.5ex]
 \hline\hline
 \rule{0pt}{10pt} MBBH & $10^6$ & $5\times10^5$ & $3 \times 10^4$ & $10^3$ & 0.4 - 600 \\
 \hline
 \rule{0pt}{10pt} IBBH & $10^4$ & $5\times10^3$ & $708$ & $10^3$ & 0.4 - 600 \\
 \hline
 \rule{0pt}{10pt} IMRI & $10^4$ & $10$ & $64$ & $10^3$ & 8 - 80 \\
 \hline
 \rule{0pt}{10pt} EMRI & $10^5$ & $10$ & $22$ & $10^3$ & 20 - 400 \\
 \hline
 \rule{0pt}{10pt} WD & $0.4$ & $0.3$ & $7$ & $5 \times 10^{-3}$ & $10^4$ \\
 \hline
\end{tabular}
\caption{Masses, SNR (Signal-to-Noise Ratio), luminosity distance and expected number of detections 
for the events that we consider for LISA (for a four-year observational time).}
\label{table:events-LISA}
\end{table}

\begin{figure}[ht]
\centering
\includegraphics[height=6.cm,width=0.5\textwidth]{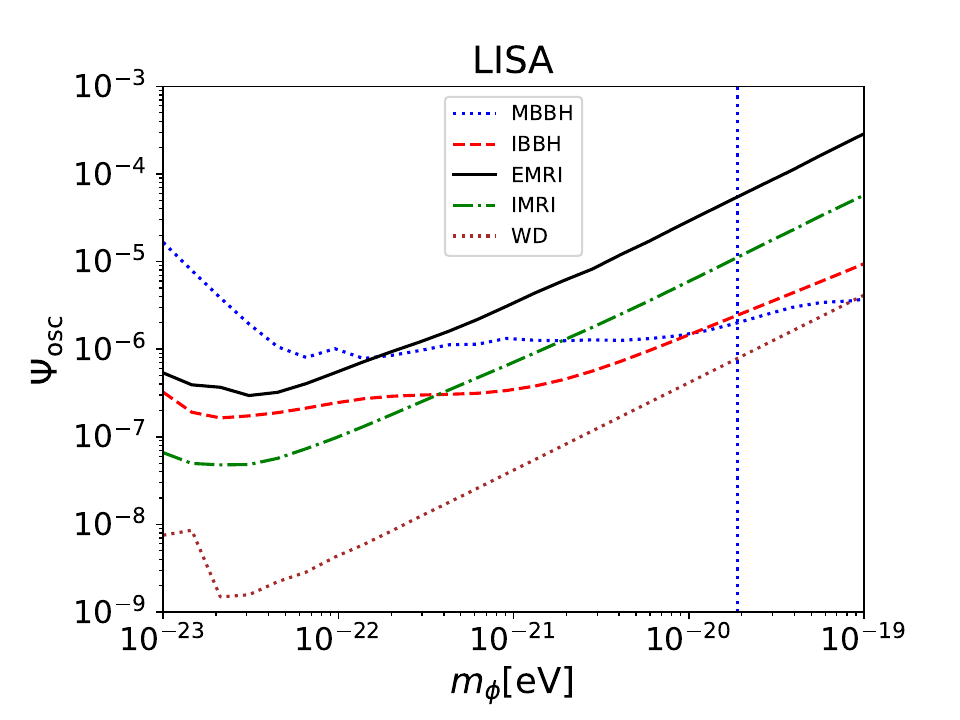}
\includegraphics[height=6.5cm,width=0.5\textwidth]{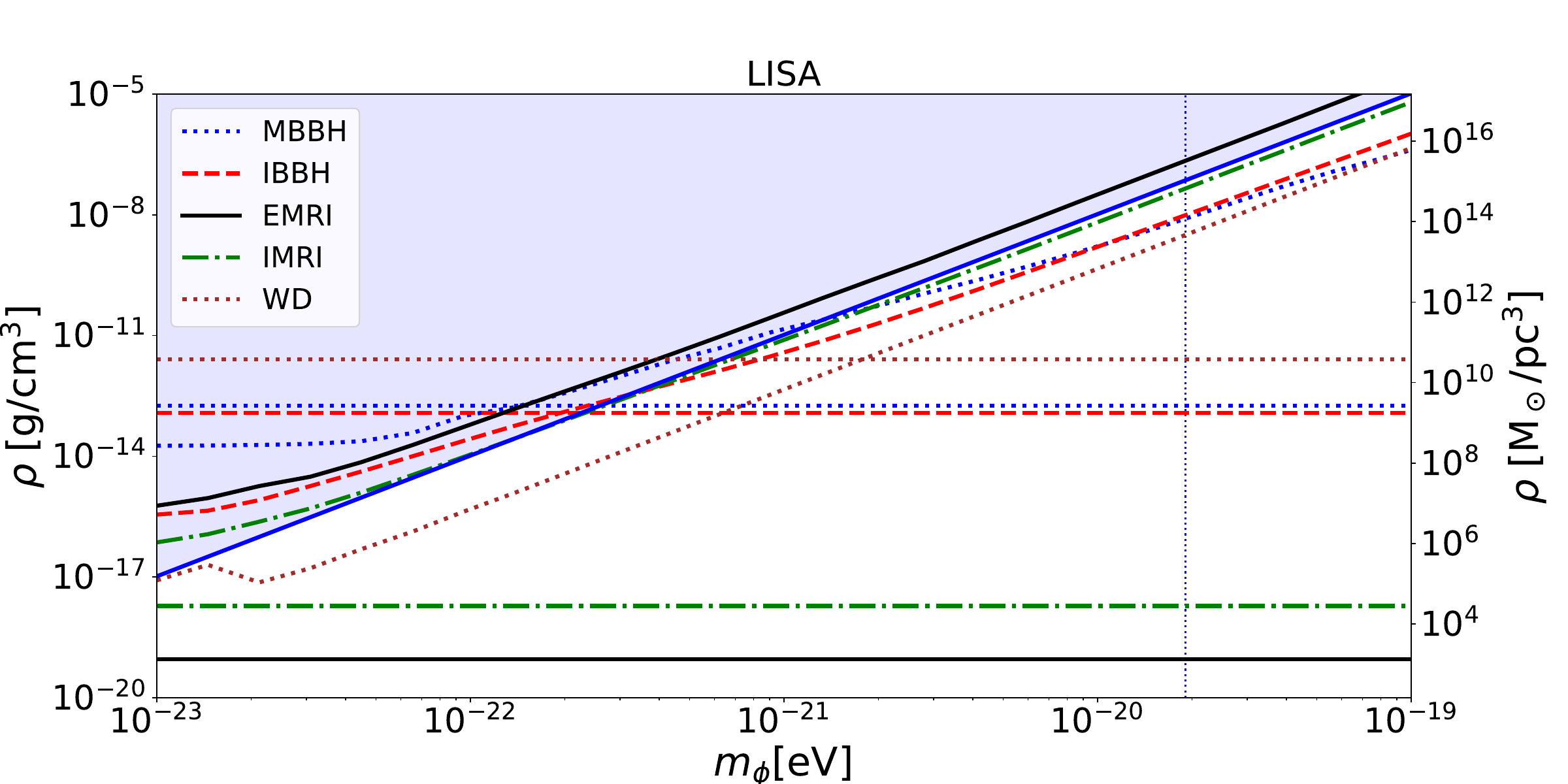}
\caption{
Detection thresholds on the amplitude $\Psi_{\rm osc}$ of the oscillating DM gravitational potential
(upper panel) and on the DM density $\rho$ (lower panel), as a function of the scalar mass $m_\phi$.
We show the results obtained for various events with the LISA interferometer.
In the lower panel, the shaded blue area is the exclusion region associated with the
upper bound (\ref{eq:rho-upper-bound}) with $M_{\rm cloud}=10^7 M_\odot$.
}
\label{fig:LISA}
\end{figure}

We now consider various binary systems that should be observed by the LISA interferometer:
Massive Binary Black Holes (MBBH), Intermediary Binary Black Holes (IBBH),
Intermediate Mass Ratio Inspirals (IMRI), Extreme Mass Ratio Inspirals (EMRI), and White Dwarfs
binaries (WD).
We give in Table~\ref{table:events-LISA} the masses, SNR, luminosity distance $d_L$ and expected
number of detections over four years of the typical events that we use for the numerical computations.
The predictions of the numbers of events involving massive BHs are very uncertain, as shown by the 
range of the estimates given in Table~\ref{table:events-LISA}, obtained from 
\cite{Klein:2015hvg,Bonetti:2018tpf,Dayal:2018gwg,Ricarte:2018}. 
In contrast, LISA is guaranteed to observed many WD binaries 
\cite{Ruiter:2007xx,Breivik:2019lmt,LISA:2022yao}.
Note that the detection thresholds obtained in Figs.~\ref{fig:LISA} and \ref{fig:DECIGO}
are for a single event and the estimates for the number of detections shown in 
Tables~\ref{table:events-LISA} and \ref{table:events-DECIGO} are only given as an indication of 
the likelihood of such events.

We show in the upper panel in Fig.~\ref{fig:LISA} our results for the detection threshold on
the oscillating DM gravitational potential $\Psi_{\rm osc}$, for these events.
The vertical blue dotted line is the upper boundary, Eq.~(\ref{eq:f-omega-opt}), for the MBBH
case. For other events this upper boundary is located to the right of the DM particle mass range
shown in the picture.
As explained in Sec.~\ref{sec:GW-phase-shift}, the phase shift of Eq.~(\ref{eq:Psi-osc-gamma})
due to the DM oscillating gravitational potential is degenerate at low and high masses with the standard result.
Thus, the amplitude $\Psi_{\rm osc}$ is poorly constrained at low and high masses and the best
constraints are obtained for $m_\phi \sim 10^{-22}$ eV.
We do not consider masses below $10^{-23}$ eV because they cannot constitute a large fraction
of the DM (the de Broglie wavelength would be greater than galactic cores).
In agreement with Eq.~(\ref{eq:M-star-def}), MBBH and IBBH events, which have a greater
chirp mass ${\cal M}$, probe somewhat higher scalar masses than EMRI and IMRI events.
White Dwarfs have smaller mass than these massive BHs. This increases the phase shift
(\ref{eq:Psi-osc-gamma-abs}) and improves the detection threshold, in agreement with our numerical
result shown in Fig.~\ref{fig:LISA}.

We show in the lower panel in Fig.~\ref{fig:LISA} our results for the detection threshold on
the DM density $\rho$.
From Eq.~(\ref{eq:Psi-osc-rho}) we have $\sigma_\rho \propto m^2 \sigma_{\Psi_{\rm osc}}$.
This leads to the very fast growth with $m_\phi$ of the detection threshold $\sigma_\rho$.
In addition to these curves, the horizontal lines show the detection thresholds
associated with the dynamical friction, Eq.~(\ref{eq:psi-df}), which are independent of $m_\phi$.
We can see that for $m_\phi \gtrsim 10^{-21}$ eV dynamical friction is more important than
the oscillatory gravitational potential. For EMRIs and IMRIs this is actually the case at all masses.
As compared with the DM density in the solar neighborhood, $\rho \sim 1 M_{\odot}/{\rm pc}^3$,
LISA can only detect DM densities that are  higher by a factor of at least $10^5$.
Such DM clouds may be formed at high redshifts, $z \sim 10^4$. This would correspond to the
matter density at the formation time when such clumps form by a rapid instability, e.g. tachyonic \cite{Brax:2020oye}.
However, their very large radii make such a scenario somewhat unlikely.
Indeed, we can expect their radius to be greater than the Compton wavelength,
\be
\lambda_C = \frac{2\pi}{m_{\phi}} = \left( \frac{m_{\phi}}{1 \, {\rm eV}} \right)^{-1}
4 \times 10^{-23} \, {\rm pc} .
\label{eq:Compton}
\ee
For $m_\phi \sim 10^{-22}$ eV, as for Fuzzy DM scenarios, this corresponds to clouds
of parsec size or greater. They would be smaller than globular clusters, which can reach sizes of  100 pc,
but denser by a factor $10^3$.
The comparison between the Compton and de Broglie wavelengths, $\lambda_{\rm dB} = \lambda_C/v$,
suggests that clouds with $R \sim \lambda_C$ would also be relativistic.
For a given mass $M_{\rm cloud}$ of the DM cloud, the inequality $R>\lambda_c$
of the cloud radius gives the upper bound
\be
\rho = \frac{M_{\rm cloud}}{R^3} < \frac{M_{\rm cloud}}{\lambda_c^3} =
\frac{M_{\rm cloud}}{1 M_\odot} \left( \frac{m_\phi}{1 \, {\rm eV}} \right)^3 \,
10^{45} \, {\rm g/cm}^3 .
\label{eq:rho-upper-bound}
\ee
We show this upper bound with $M_{\rm cloud}=10^7 M_\odot$ by the blue shaded area in the lower panel
in Fig.~\ref{fig:LISA}. Thus, we can see that the high densities required to detect the phase shift
$\Delta\psi_{\rm osc}$ also imply very high cloud masses, $M_{\rm cloud} \gtrsim 10^5 M_\odot$
for WD binaries.

\subsection{DECIGO}

\begin{table}[hbtp]
\centering
\begin{tabular}{|l||*{5}{c|}}
 \hline
 \rule{0pt}{10pt} & $m_1$ ($\rm M_{\odot}$) & $m_2$ ($\rm M_{\odot}$) & SNR  & $d_L$ (Mpc) & detections \\ [0.5ex]
 \hline\hline
 \rule{0pt}{10pt} GW150914 & $35.6$ & $30.6$ & $2815$ & 440 &  $ > 10^4$ \\
 \hline
 \rule{0pt}{10pt} GW170608 & $11$ & $7.6$ & $1290$ & 320 & $ > 10^4$ \\
 \hline
 \rule{0pt}{10pt} GW170817 & $1.46$ & $1.27$ & $2124$ & 40 & $ 10^5 $ \\
 \hline
 \rule{0pt}{10pt} WD & $0.4$ & $0.3$ & $8$ & 375 & $ > 6600 $ \\
 \hline
 \end{tabular}
\caption{Masses, SNR, luminosity distance and expected number of detections of the events that we consider for DECIGO, for a one-year observational time.}
\label{table:events-DECIGO}
\end{table}

We also consider stellar-mass BHs,  neutron stars and white dwarfs events that could be detected 
by the DECIGO interferometer. We choose as typical cases three events detected by LIGO and Virgo
\cite{LIGOScientific:2018mvr} given in Table~\ref{table:events-DECIGO}, as well as a typical white dwarf 
merger. The expected detection rates are obtained from \cite{Seto-2001,Nakamura-2016,Kawamura:2020pcg,Kinugawa2022}.
We can see that many events are expected for these four classes of binaries.

\begin{figure}[ht]
\centering
\includegraphics[height=6.cm,width=0.5\textwidth]{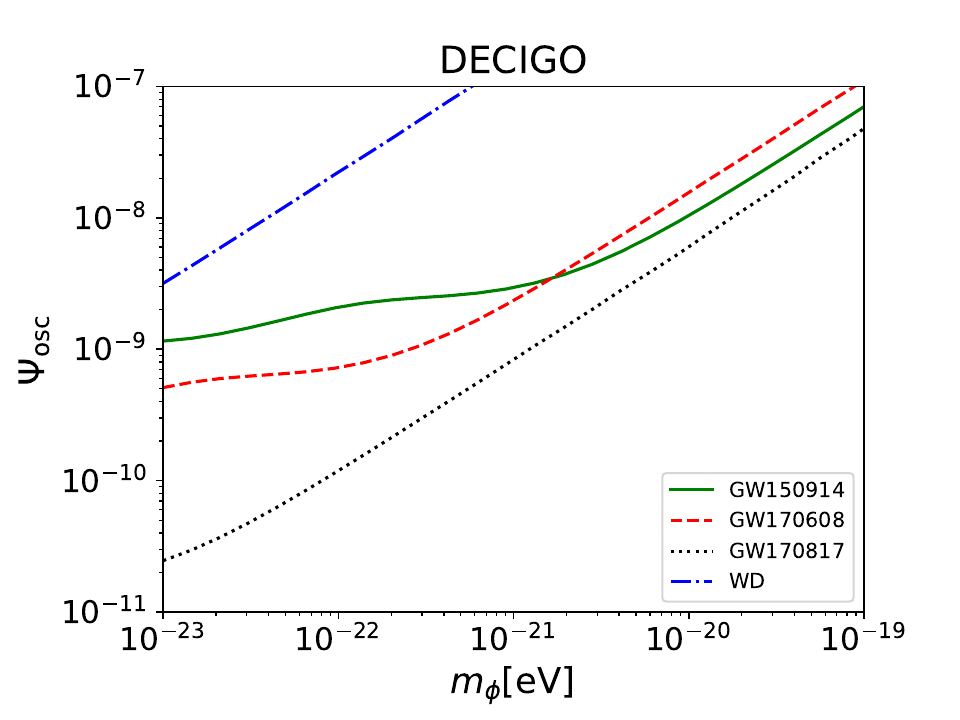}
\includegraphics[height=6.5cm,width=0.5\textwidth]{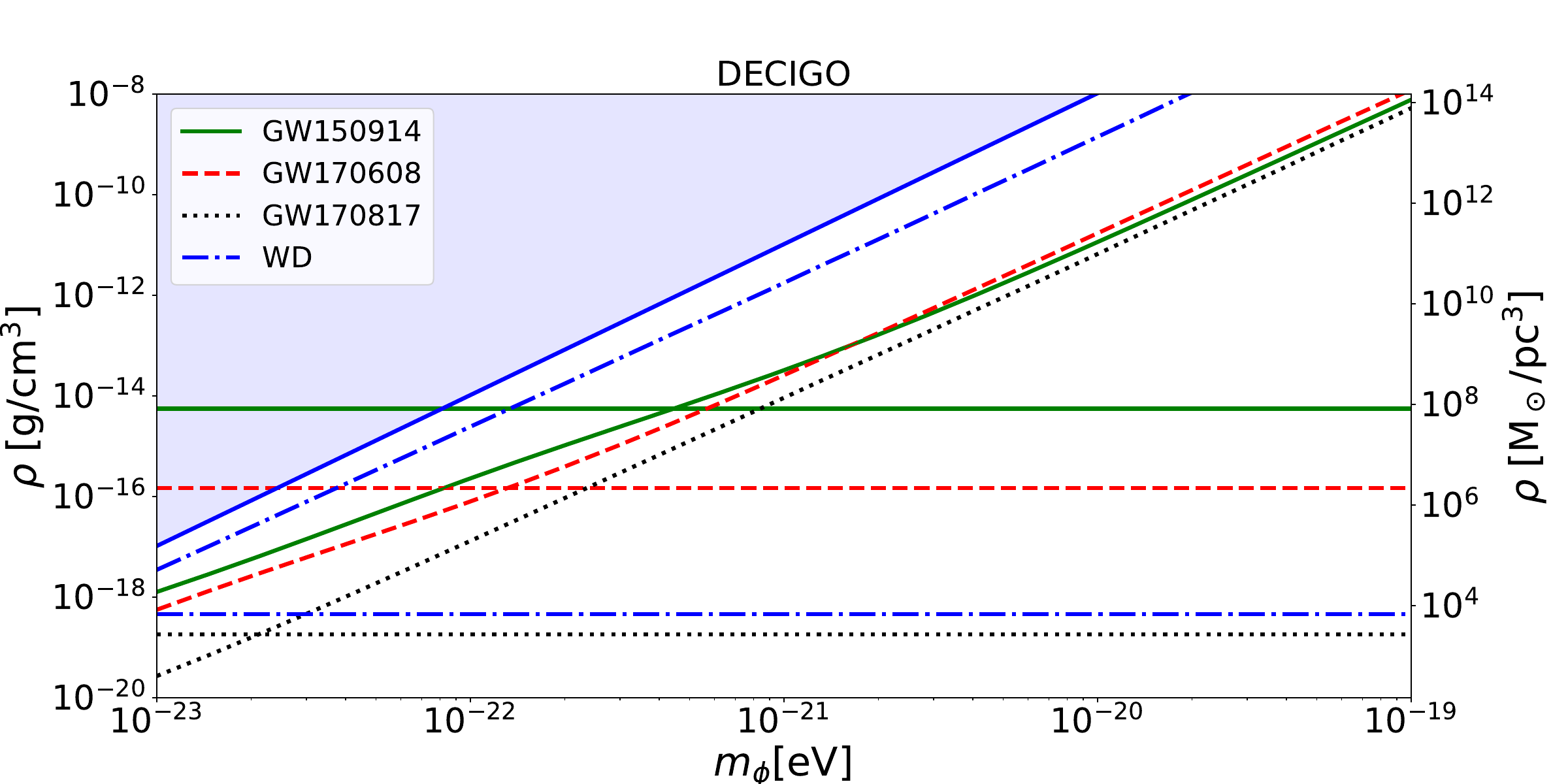}
\caption{
DM detection thresholds as in Fig.~\ref{fig:LISA} but for the DECIGO interferometer.
}
\label{fig:DECIGO}
\end{figure}

We show in Fig.~\ref{fig:DECIGO} our results for the detection thresholds on the oscillating DM
gravitational potential $\Psi_{\rm osc}$ and the density $\rho$, for various events with the DECIGO
interferometer.
For all events the upper boundary, Eq.~(\ref{eq:f-omega-opt}), is located to the right of the
DM particle mass range shown in the picture.
The thresholds for DECIGO and LISA are of about the same orders of magnitude, although they are
somewhat more favorable for DECIGO. In particular, the required DM density are
further below the upper bound of Eq.~(\ref{eq:rho-upper-bound}).
For NS and WD binaries the GW waveform is more sensitive to dynamical friction than to the DM oscillations
for almost all DM masses. For BH binaries the signal associated with the DM oscillations dominates over dynamical friction for $m_\phi \lesssim 10^{-21}$ eV.

\section{conclusions}
\label{sec:conc}

In this work, we have examined whether the oscillatory behavior of the gravitational potential
of DM halos predicted by some DM scenarios could be detected by gravitational
wave interferometers such as LISA and DECIGO, if  binary systems were embedded within
high-density DM clouds.
Building on the early work in Ref.~\cite{Khmelnitsky:2013lxt}, which considered the impact of these
oscillations on pulsar timing signals, we now consider  their impact on the phase of the
GW form received by interferometers.
We derived the associated phase shift and performed a Fisher analysis to estimate the detection
thresholds that can be expected for near future instruments, for a variety of binary systems.

We find that this probe is unlikely to be competitive with more direct observations of DM
substructures. For $m_\phi > 10^{-21} \, {\rm eV}$ the effect of the DM environment
on the GW form due to the usual dynamical friction (the drag force
that contributes to the shrinking of the orbital radius of the binary system) is expected to dominate
over the effect associated with these oscillatory features of the DM gravitational potential
(which only affect a subleading component of $\Psi_N$).
For low particle masses below $10^{-23} \, {\rm eV}$, the scalar clouds are associated with
Compton wavelengths greater than the parsec scale. This implies DM clouds that
are too large to provide realistic DM scenarios.

For DM masses $m_\phi \sim 10^{-22} \, {\rm eV}$, the phase shift associated with the
oscillations of the DM gravitational potential can only be detected by LISA or DECIGO
for densities that are greater than that in the solar neighborhood by a factor $10^5$ (LISA)
or $10^4$ (DECIGO). This would also correspond to cloud masses above $10^5 M_\odot$ (LISA)
or $10^3 M_\odot$ (DECIGO) and radii above $0.4$ pc.
Although such high-density structures may be possible, if they formed at redshifts
$ z \sim 10^4$, this would require a non-standard formation mechanism,
such as instabilities due to DM self-interactions. In this sense, LISA and DECIGO would only be sensitive
to the oscillatory features from exotic types of DM.

Therefore, except for a small region in the parameter space of DM models,
the phase of the GW wave form is unlikely to be sensitive to the oscillatory features
of DM gravitational potentials.
This justifies standard analyses of the emission of GWs by binary systems,
where the DM environment is neglected or considered through its usual effects:
dynamical friction, accretion and gravitational pull by the enclosed DM mass within the orbital
radius. On the other hand, from a beyond the standard model perspective, LISA and DECIGO could
provide us with a window on the physics of dark matter and its possible exotic properties in the radiation era before large scale structures of the Universe form.

\vspace{1cm}
\section*{Acknowledgements}

This work was partially supported by the MICINN (Ministerio de Ciencia e Innovación, Spain) projects  PID2019-107394GB-I00/AEI/10.13039/501100011033 (AEI/FEDER, UE) and PID2022-139841NB-I00, COST (European Cooperation in Science and Technology) Actions CA21106 and CA21136. J.A.R.C. acknowledges support by Institut Pascal at Université Paris-Saclay during the Paris-Saclay Astroparticle Symposium 2022, with the support of the P2IO Laboratory of Excellence (program “Investissements d’avenir” ANR-11-IDEX-0003-01 Paris-Saclay and ANR-10-LABX-0038), the P2I axis of the Graduate School of Physics of Université Paris-Saclay, as well as IJCLab, CEA, APPEC, IAS, OSUPS, and the IN2P3 master project UCMN. C.B. is supported by the STFC under grant ST/T000732/1.

\bibliography{GW}

\begin{thebibliography}{67}
\expandafter\ifx\csname natexlab\endcsname\relax\def\natexlab#1{#1}\fi
\expandafter\ifx\csname bibnamefont\endcsname\relax
  \def\bibnamefont#1{#1}\fi
\expandafter\ifx\csname bibfnamefont\endcsname\relax
  \def\bibfnamefont#1{#1}\fi
\expandafter\ifx\csname citenamefont\endcsname\relax
  \def\citenamefont#1{#1}\fi
\expandafter\ifx\csname url\endcsname\relax
  \def\url#1{\texttt{#1}}\fi
\expandafter\ifx\csname urlprefix\endcsname\relax\def\urlprefix{URL }\fi
\providecommand{\bibinfo}[2]{#2}
\providecommand{\eprint}[2][]{\url{#2}}

\bibitem[{\citenamefont{Hui et~al.}(2017)\citenamefont{Hui, Ostriker, Tremaine,
  and Witten}}]{Hui:2016ltb}
\bibinfo{author}{\bibfnamefont{L.}~\bibnamefont{Hui}},
  \bibinfo{author}{\bibfnamefont{J.~P.} \bibnamefont{Ostriker}},
  \bibinfo{author}{\bibfnamefont{S.}~\bibnamefont{Tremaine}}, \bibnamefont{and}
  \bibinfo{author}{\bibfnamefont{E.}~\bibnamefont{Witten}},
  \bibinfo{journal}{Phys. Rev. D} \textbf{\bibinfo{volume}{95}},
  \bibinfo{pages}{043541} (\bibinfo{year}{2017}), \eprint{1610.08297}.

\bibitem[{\citenamefont{Bar et~al.}(2018)\citenamefont{Bar, Blas, Blum, and
  Sibiryakov}}]{Bar:2018acw}
\bibinfo{author}{\bibfnamefont{N.}~\bibnamefont{Bar}},
  \bibinfo{author}{\bibfnamefont{D.}~\bibnamefont{Blas}},
  \bibinfo{author}{\bibfnamefont{K.}~\bibnamefont{Blum}}, \bibnamefont{and}
  \bibinfo{author}{\bibfnamefont{S.}~\bibnamefont{Sibiryakov}},
  \bibinfo{journal}{Phys. Rev. D} \textbf{\bibinfo{volume}{98}},
  \bibinfo{pages}{083027} (\bibinfo{year}{2018}), \eprint{1805.00122}.

\bibitem[{\citenamefont{Cembranos et~al.}(2016)\citenamefont{Cembranos, Maroto,
  and N\'u\~nez Jare\~no}}]{Cembranos:2015oya}
\bibinfo{author}{\bibfnamefont{J.~A.~R.} \bibnamefont{Cembranos}},
  \bibinfo{author}{\bibfnamefont{A.~L.} \bibnamefont{Maroto}},
  \bibnamefont{and} \bibinfo{author}{\bibfnamefont{S.~J.}
  \bibnamefont{N\'u\~nez Jare\~no}}, \bibinfo{journal}{JHEP}
  \textbf{\bibinfo{volume}{03}}, \bibinfo{pages}{013} (\bibinfo{year}{2016}),
  \eprint{1509.08819}.

\bibitem[{\citenamefont{Turner}(1983)}]{Turner:1983he}
\bibinfo{author}{\bibfnamefont{M.~S.} \bibnamefont{Turner}},
  \bibinfo{journal}{Phys. Rev. D} \textbf{\bibinfo{volume}{28}},
  \bibinfo{pages}{1243} (\bibinfo{year}{1983}).

\bibitem[{\citenamefont{Johnson and Kamionkowski}(2008)}]{Johnson:2008se}
\bibinfo{author}{\bibfnamefont{M.~C.} \bibnamefont{Johnson}} \bibnamefont{and}
  \bibinfo{author}{\bibfnamefont{M.}~\bibnamefont{Kamionkowski}},
  \bibinfo{journal}{Phys. Rev. D} \textbf{\bibinfo{volume}{78}},
  \bibinfo{pages}{063010} (\bibinfo{year}{2008}), \eprint{0805.1748}.

\bibitem[{\citenamefont{Brax et~al.}(2019)\citenamefont{Brax, Cembranos, and
  Valageas}}]{Brax:2019fzb}
\bibinfo{author}{\bibfnamefont{P.}~\bibnamefont{Brax}},
  \bibinfo{author}{\bibfnamefont{J.~A.~R.} \bibnamefont{Cembranos}},
  \bibnamefont{and} \bibinfo{author}{\bibfnamefont{P.}~\bibnamefont{Valageas}},
  \bibinfo{journal}{Phys. Rev. D} \textbf{\bibinfo{volume}{100}},
  \bibinfo{pages}{023526} (\bibinfo{year}{2019}), \eprint{1906.00730}.

\bibitem[{\citenamefont{Hu et~al.}(2000)\citenamefont{Hu, Barkana, and
  Gruzinov}}]{Hu2000}
\bibinfo{author}{\bibfnamefont{W.}~\bibnamefont{Hu}},
  \bibinfo{author}{\bibfnamefont{R.}~\bibnamefont{Barkana}}, \bibnamefont{and}
  \bibinfo{author}{\bibfnamefont{A.}~\bibnamefont{Gruzinov}},
  \bibinfo{journal}{Physical Review Letters} \textbf{\bibinfo{volume}{85}},
  \bibinfo{pages}{1158} (\bibinfo{year}{2000}), ISSN \bibinfo{issn}{00319007}.

\bibitem[{\citenamefont{Schive et~al.}(2014)\citenamefont{Schive, Chiueh, and
  Broadhurst}}]{Schive2014}
\bibinfo{author}{\bibfnamefont{H.~Y.} \bibnamefont{Schive}},
  \bibinfo{author}{\bibfnamefont{T.}~\bibnamefont{Chiueh}}, \bibnamefont{and}
  \bibinfo{author}{\bibfnamefont{T.}~\bibnamefont{Broadhurst}},
  \bibinfo{journal}{Nature Physics} \textbf{\bibinfo{volume}{10}},
  \bibinfo{pages}{496} (\bibinfo{year}{2014}), ISSN \bibinfo{issn}{17452481}.

\bibitem[{\citenamefont{Mocz et~al.}(2017)\citenamefont{Mocz, Vogelsberger,
  Robles, Zavala, Boylan-Kolchin, Fialkov, and Hernquist}}]{Mocz2017}
\bibinfo{author}{\bibfnamefont{P.}~\bibnamefont{Mocz}},
  \bibinfo{author}{\bibfnamefont{M.}~\bibnamefont{Vogelsberger}},
  \bibinfo{author}{\bibfnamefont{V.~H.} \bibnamefont{Robles}},
  \bibinfo{author}{\bibfnamefont{J.}~\bibnamefont{Zavala}},
  \bibinfo{author}{\bibfnamefont{M.}~\bibnamefont{Boylan-Kolchin}},
  \bibinfo{author}{\bibfnamefont{A.}~\bibnamefont{Fialkov}}, \bibnamefont{and}
  \bibinfo{author}{\bibfnamefont{L.}~\bibnamefont{Hernquist}},
  \bibinfo{journal}{Monthly Notices of the Royal Astronomical Society}
  \textbf{\bibinfo{volume}{471}}, \bibinfo{pages}{4559} (\bibinfo{year}{2017}),
  ISSN \bibinfo{issn}{13652966}.

\bibitem[{\citenamefont{May and Springel}(2021)}]{May2021StructureDynamics}
\bibinfo{author}{\bibfnamefont{S.}~\bibnamefont{May}} \bibnamefont{and}
  \bibinfo{author}{\bibfnamefont{V.}~\bibnamefont{Springel}},
  \bibinfo{journal}{Monthly Notices of the Royal Astronomical Society}
  \textbf{\bibinfo{volume}{506}}, \bibinfo{pages}{2603} (\bibinfo{year}{2021}),
  ISSN \bibinfo{issn}{0035-8711},
  \urlprefix\url{https://academic.oup.com/mnras/article/506/2/2603/6308377}.

\bibitem[{\citenamefont{Macci{\`{o}} et~al.}(2012)\citenamefont{Macci{\`{o}},
  Paduroiu, Anderhalden, Schneider, and Moore}}]{Maccio1}
\bibinfo{author}{\bibfnamefont{A.~V.} \bibnamefont{Macci{\`{o}}}},
  \bibinfo{author}{\bibfnamefont{S.}~\bibnamefont{Paduroiu}},
  \bibinfo{author}{\bibfnamefont{D.}~\bibnamefont{Anderhalden}},
  \bibinfo{author}{\bibfnamefont{A.}~\bibnamefont{Schneider}},
  \bibnamefont{and} \bibinfo{author}{\bibfnamefont{B.}~\bibnamefont{Moore}},
  \bibinfo{journal}{Monthly Notices of the Royal Astronomical Society}
  \textbf{\bibinfo{volume}{424}}, \bibinfo{pages}{1105} (\bibinfo{year}{2012}),
  ISSN \bibinfo{issn}{13652966},
  \urlprefix\url{https://academic.oup.com/mnras/article/424/2/1105/998349}.

\bibitem[{\citenamefont{{Weinberg} et~al.}(2015)\citenamefont{{Weinberg},
  {Bullock}, {Governato}, {Kuzio de Naray}, and {Peter}}}]{2015PNAS..11212249W}
\bibinfo{author}{\bibfnamefont{D.~H.} \bibnamefont{{Weinberg}}},
  \bibinfo{author}{\bibfnamefont{J.~S.} \bibnamefont{{Bullock}}},
  \bibinfo{author}{\bibfnamefont{F.}~\bibnamefont{{Governato}}},
  \bibinfo{author}{\bibfnamefont{R.}~\bibnamefont{{Kuzio de Naray}}},
  \bibnamefont{and} \bibinfo{author}{\bibfnamefont{A.~H.~G.}
  \bibnamefont{{Peter}}}, \bibinfo{journal}{Proceedings of the National Academy
  of Science} \textbf{\bibinfo{volume}{112}}, \bibinfo{pages}{12249}
  (\bibinfo{year}{2015}).

\bibitem[{\citenamefont{{Bullock} and
  {Boylan-Kolchin}}(2017)}]{2017ARA&A..55..343B}
\bibinfo{author}{\bibfnamefont{J.~S.} \bibnamefont{{Bullock}}}
  \bibnamefont{and}
  \bibinfo{author}{\bibfnamefont{M.}~\bibnamefont{{Boylan-Kolchin}}},
  \bibinfo{journal}{Annual Review of Astronomy and Astrophysics}
  \textbf{\bibinfo{volume}{55}}, \bibinfo{pages}{343} (\bibinfo{year}{2017}).

\bibitem[{\citenamefont{{Del Popolo} and {Le
  Delliou}}(2017)}]{2017Galax...5...17D}
\bibinfo{author}{\bibfnamefont{A.}~\bibnamefont{{Del Popolo}}}
  \bibnamefont{and} \bibinfo{author}{\bibfnamefont{M.}~\bibnamefont{{Le
  Delliou}}}, \bibinfo{journal}{Galaxies} \textbf{\bibinfo{volume}{5}},
  \bibinfo{pages}{17} (\bibinfo{year}{2017}).

\bibitem[{\citenamefont{Mocz et~al.}(2019)\citenamefont{Mocz, Fialkov,
  Vogelsberger, Becerra, Amin, Bose, Boylan-Kolchin, Chavanis, Hernquist,
  Lancaster et~al.}}]{Mocz2019}
\bibinfo{author}{\bibfnamefont{P.}~\bibnamefont{Mocz}},
  \bibinfo{author}{\bibfnamefont{A.}~\bibnamefont{Fialkov}},
  \bibinfo{author}{\bibfnamefont{M.}~\bibnamefont{Vogelsberger}},
  \bibinfo{author}{\bibfnamefont{F.}~\bibnamefont{Becerra}},
  \bibinfo{author}{\bibfnamefont{M.~A.} \bibnamefont{Amin}},
  \bibinfo{author}{\bibfnamefont{S.}~\bibnamefont{Bose}},
  \bibinfo{author}{\bibfnamefont{M.}~\bibnamefont{Boylan-Kolchin}},
  \bibinfo{author}{\bibfnamefont{P.~H.} \bibnamefont{Chavanis}},
  \bibinfo{author}{\bibfnamefont{L.}~\bibnamefont{Hernquist}},
  \bibinfo{author}{\bibfnamefont{L.}~\bibnamefont{Lancaster}},
  \bibnamefont{et~al.}, \bibinfo{journal}{Physical Review Letters}
  \textbf{\bibinfo{volume}{123}}, \bibinfo{pages}{141301}
  (\bibinfo{year}{2019}), ISSN \bibinfo{issn}{10797114},
  \urlprefix\url{https://journals.aps.org/prl/abstract/10.1103/PhysRevLett.123.141301}.

\bibitem[{\citenamefont{Mocz et~al.}(2020)\citenamefont{Mocz, Fialkov,
  Vogelsberger, Becerra, Shen, Robles, Amin, Zavala, Boylan-Kolchin, Bose
  et~al.}}]{Mocz2020}
\bibinfo{author}{\bibfnamefont{P.}~\bibnamefont{Mocz}},
  \bibinfo{author}{\bibfnamefont{A.}~\bibnamefont{Fialkov}},
  \bibinfo{author}{\bibfnamefont{M.}~\bibnamefont{Vogelsberger}},
  \bibinfo{author}{\bibfnamefont{F.}~\bibnamefont{Becerra}},
  \bibinfo{author}{\bibfnamefont{X.}~\bibnamefont{Shen}},
  \bibinfo{author}{\bibfnamefont{V.~H.} \bibnamefont{Robles}},
  \bibinfo{author}{\bibfnamefont{M.~A.} \bibnamefont{Amin}},
  \bibinfo{author}{\bibfnamefont{J.}~\bibnamefont{Zavala}},
  \bibinfo{author}{\bibfnamefont{M.}~\bibnamefont{Boylan-Kolchin}},
  \bibinfo{author}{\bibfnamefont{S.}~\bibnamefont{Bose}}, \bibnamefont{et~al.},
  \bibinfo{journal}{Monthly Notices of the Royal Astronomical Society}
  \textbf{\bibinfo{volume}{494}}, \bibinfo{pages}{2027} (\bibinfo{year}{2020}),
  ISSN \bibinfo{issn}{0035-8711},
  \urlprefix\url{https://academic.oup.com/mnras/article/494/2/2027/5819969}.

\bibitem[{\citenamefont{Banerjee et~al.}(2020)\citenamefont{Banerjee, Bera, and
  Mota}}]{Banerjee2020}
\bibinfo{author}{\bibfnamefont{S.}~\bibnamefont{Banerjee}},
  \bibinfo{author}{\bibfnamefont{S.}~\bibnamefont{Bera}}, \bibnamefont{and}
  \bibinfo{author}{\bibfnamefont{D.~F.} \bibnamefont{Mota}},
  \bibinfo{journal}{Journal of Cosmology and Astroparticle Physics}
  \textbf{\bibinfo{volume}{2020}}, \bibinfo{pages}{034} (\bibinfo{year}{2020}),
  ISSN \bibinfo{issn}{1475-7516},
  \urlprefix\url{https://iopscience.iop.org/article/10.1088/1475-7516/2020/07/034
  https://iopscience.iop.org/article/10.1088/1475-7516/2020/07/034/meta}.

\bibitem[{\citenamefont{Mina et~al.}(2022)\citenamefont{Mina, Mota, and
  Winther}}]{Mina2022}
\bibinfo{author}{\bibfnamefont{M.}~\bibnamefont{Mina}},
  \bibinfo{author}{\bibfnamefont{D.~F.} \bibnamefont{Mota}}, \bibnamefont{and}
  \bibinfo{author}{\bibfnamefont{H.~A.} \bibnamefont{Winther}},
  \bibinfo{journal}{Astronomy {\&} Astrophysics}
  \textbf{\bibinfo{volume}{662}}, \bibinfo{pages}{A29} (\bibinfo{year}{2022}),
  ISSN \bibinfo{issn}{0004-6361},
  \urlprefix\url{https://www.aanda.org/articles/aa/full_html/2022/06/aa38876-20/aa38876-20.html
  https://www.aanda.org/articles/aa/abs/2022/06/aa38876-20/aa38876-20.html}.

\bibitem[{\citenamefont{Li et~al.}(2021)\citenamefont{Li, Hui, and
  Yavetz}}]{Li2021}
\bibinfo{author}{\bibfnamefont{X.}~\bibnamefont{Li}},
  \bibinfo{author}{\bibfnamefont{L.}~\bibnamefont{Hui}}, \bibnamefont{and}
  \bibinfo{author}{\bibfnamefont{T.~D.} \bibnamefont{Yavetz}},
  \bibinfo{journal}{Physical Review D} \textbf{\bibinfo{volume}{103}}
  (\bibinfo{year}{2021}), ISSN \bibinfo{issn}{24700029}.

\bibitem[{\citenamefont{Hui}(2021)}]{Hui2021-2}
\bibinfo{author}{\bibfnamefont{L.}~\bibnamefont{Hui}}, \bibinfo{journal}{Annual
  Review of Astronomy and Astrophysics} \textbf{\bibinfo{volume}{59}},
  \bibinfo{pages}{247} (\bibinfo{year}{2021}), ISSN \bibinfo{issn}{00664146},
  \urlprefix\url{https://www.annualreviews.org/doi/abs/10.1146/annurev-astro-120920-010024}.

\bibitem[{\citenamefont{Woo and Chiueh}(2009)}]{Woo1}
\bibinfo{author}{\bibfnamefont{T.~P.} \bibnamefont{Woo}} \bibnamefont{and}
  \bibinfo{author}{\bibfnamefont{T.}~\bibnamefont{Chiueh}},
  \bibinfo{journal}{The Astrophysical Journal} \textbf{\bibinfo{volume}{697}},
  \bibinfo{pages}{850} (\bibinfo{year}{2009}), ISSN \bibinfo{issn}{0004-637X},
  \urlprefix\url{https://iopscience.iop.org/article/10.1088/0004-637X/697/1/850
  https://iopscience.iop.org/article/10.1088/0004-637X/697/1/850/meta}.

\bibitem[{\citenamefont{Marsh and Niemeyer}(2019)}]{Marsh2019}
\bibinfo{author}{\bibfnamefont{D.~J.} \bibnamefont{Marsh}} \bibnamefont{and}
  \bibinfo{author}{\bibfnamefont{J.~C.} \bibnamefont{Niemeyer}},
  \bibinfo{journal}{Physical Review Letters} \textbf{\bibinfo{volume}{123}}
  (\bibinfo{year}{2019}), ISSN \bibinfo{issn}{10797114}.

\bibitem[{\citenamefont{Veltmaat et~al.}(2018)\citenamefont{Veltmaat, Niemeyer,
  and Schwabe}}]{Veltmaat2018}
\bibinfo{author}{\bibfnamefont{J.}~\bibnamefont{Veltmaat}},
  \bibinfo{author}{\bibfnamefont{J.~C.} \bibnamefont{Niemeyer}},
  \bibnamefont{and} \bibinfo{author}{\bibfnamefont{B.}~\bibnamefont{Schwabe}},
  \bibinfo{journal}{Physical Review D} \textbf{\bibinfo{volume}{98}}
  (\bibinfo{year}{2018}), ISSN \bibinfo{issn}{24700029}.

\bibitem[{\citenamefont{Glennon and Prescod-Weinstein}(2021)}]{Glennon2020-2}
\bibinfo{author}{\bibfnamefont{N.}~\bibnamefont{Glennon}} \bibnamefont{and}
  \bibinfo{author}{\bibfnamefont{C.}~\bibnamefont{Prescod-Weinstein}},
  \bibinfo{journal}{Physical Review D} \textbf{\bibinfo{volume}{104}},
  \bibinfo{pages}{083532} (\bibinfo{year}{2021}), ISSN
  \bibinfo{issn}{24700029},
  \urlprefix\url{https://journals.aps.org/prd/abstract/10.1103/PhysRevD.104.083532}.

\bibitem[{\citenamefont{Chavanis}(2011)}]{Chavanis2011}
\bibinfo{author}{\bibfnamefont{P.-H.} \bibnamefont{Chavanis}},
  \bibinfo{journal}{Physical Review D} \textbf{\bibinfo{volume}{84}},
  \bibinfo{pages}{043531} (\bibinfo{year}{2011}), ISSN
  \bibinfo{issn}{15507998}.

\bibitem[{\citenamefont{Chavanis and Delfini}(2011)}]{Chavanis2011_2}
\bibinfo{author}{\bibfnamefont{P.-H.} \bibnamefont{Chavanis}} \bibnamefont{and}
  \bibinfo{author}{\bibfnamefont{L.}~\bibnamefont{Delfini}},
  \bibinfo{journal}{Physical Review D} \textbf{\bibinfo{volume}{84}},
  \bibinfo{pages}{043532} (\bibinfo{year}{2011}), ISSN
  \bibinfo{issn}{1550-7998},
  \urlprefix\url{https://journals.aps.org/prd/abstract/10.1103/PhysRevD.84.043532}.

\bibitem[{\citenamefont{Rindler-Daller and
  Shapiro}(2012)}]{Rindler-Daller:2011afd}
\bibinfo{author}{\bibfnamefont{T.}~\bibnamefont{Rindler-Daller}}
  \bibnamefont{and} \bibinfo{author}{\bibfnamefont{P.~R.}
  \bibnamefont{Shapiro}}, \bibinfo{journal}{Monthly Notices of the Royal
  Astronomical Society} \textbf{\bibinfo{volume}{422}}, \bibinfo{pages}{135}
  (\bibinfo{year}{2012}).

\bibitem[{\citenamefont{Li et~al.}(2014)\citenamefont{Li, Rindler-Daller, and
  Shapiro}}]{Li2014}
\bibinfo{author}{\bibfnamefont{B.}~\bibnamefont{Li}},
  \bibinfo{author}{\bibfnamefont{T.}~\bibnamefont{Rindler-Daller}},
  \bibnamefont{and} \bibinfo{author}{\bibfnamefont{P.~R.}
  \bibnamefont{Shapiro}}, \bibinfo{journal}{Physical Review D - Particles,
  Fields, Gravitation and Cosmology} \textbf{\bibinfo{volume}{89}},
  \bibinfo{pages}{083536} (\bibinfo{year}{2014}), ISSN
  \bibinfo{issn}{15502368},
  \urlprefix\url{https://journals.aps.org/prd/abstract/10.1103/PhysRevD.89.083536}.

\bibitem[{\citenamefont{Li et~al.}(2017)\citenamefont{Li, Shapiro, and
  Rindler-Daller}}]{Li2017}
\bibinfo{author}{\bibfnamefont{B.}~\bibnamefont{Li}},
  \bibinfo{author}{\bibfnamefont{P.~R.} \bibnamefont{Shapiro}},
  \bibnamefont{and}
  \bibinfo{author}{\bibfnamefont{T.}~\bibnamefont{Rindler-Daller}},
  \bibinfo{journal}{Physical Review D} \textbf{\bibinfo{volume}{96}},
  \bibinfo{pages}{063505} (\bibinfo{year}{2017}), ISSN
  \bibinfo{issn}{24700029},
  \urlprefix\url{https://journals.aps.org/prd/abstract/10.1103/PhysRevD.96.063505}.

\bibitem[{\citenamefont{Desjacques et~al.}(2018)\citenamefont{Desjacques,
  Kehagias, and Riotto}}]{Desjacques:2017fmf}
\bibinfo{author}{\bibfnamefont{V.}~\bibnamefont{Desjacques}},
  \bibinfo{author}{\bibfnamefont{A.}~\bibnamefont{Kehagias}}, \bibnamefont{and}
  \bibinfo{author}{\bibfnamefont{A.}~\bibnamefont{Riotto}},
  \bibinfo{journal}{Physical Review D} \textbf{\bibinfo{volume}{97}},
  \bibinfo{pages}{023529} (\bibinfo{year}{2018}).

\bibitem[{\citenamefont{Chavanis}(2021)}]{Chavanis:2020rdo}
\bibinfo{author}{\bibfnamefont{P.-H.} \bibnamefont{Chavanis}},
  \bibinfo{journal}{Physical Review D} \textbf{\bibinfo{volume}{103}},
  \bibinfo{pages}{123551} (\bibinfo{year}{2021}).

\bibitem[{\citenamefont{Hartman
  et~al.}(2022{\natexlab{a}})\citenamefont{Hartman, Winther, and
  Mota}}]{Hartman2022}
\bibinfo{author}{\bibfnamefont{S.~T.} \bibnamefont{Hartman}},
  \bibinfo{author}{\bibfnamefont{H.~A.} \bibnamefont{Winther}},
  \bibnamefont{and} \bibinfo{author}{\bibfnamefont{D.~F.} \bibnamefont{Mota}},
  \bibinfo{journal}{Journal of Cosmology and Astroparticle Physics}
  \textbf{\bibinfo{volume}{2022}}, \bibinfo{pages}{005}
  (\bibinfo{year}{2022}{\natexlab{a}}), ISSN \bibinfo{issn}{1475-7516},
  \urlprefix\url{https://iopscience.iop.org/article/10.1088/1475-7516/2022/02/005
  https://iopscience.iop.org/article/10.1088/1475-7516/2022/02/005/meta}.

\bibitem[{\citenamefont{Hartman
  et~al.}(2022{\natexlab{b}})\citenamefont{Hartman, Winther, and
  Mota}}]{Hartman2022-2}
\bibinfo{author}{\bibfnamefont{S.~T.~H.} \bibnamefont{Hartman}},
  \bibinfo{author}{\bibfnamefont{H.~A.} \bibnamefont{Winther}},
  \bibnamefont{and} \bibinfo{author}{\bibfnamefont{D.~F.} \bibnamefont{Mota}},
  \bibinfo{journal}{Astronomy {\&} Astrophysics}
  \textbf{\bibinfo{volume}{666}}, \bibinfo{pages}{A95}
  (\bibinfo{year}{2022}{\natexlab{b}}), ISSN \bibinfo{issn}{0004-6361},
  \urlprefix\url{https://www.aanda.org/articles/aa/full_html/2022/10/aa43496-22/aa43496-22.html
  https://www.aanda.org/articles/aa/abs/2022/10/aa43496-22/aa43496-22.html}.

\bibitem[{\citenamefont{Mocz et~al.}(2023)\citenamefont{Mocz, Fialkov,
  Vogelsberger, Boylan-Kolchin, Chavanis, Amin, Bose, Dome, Hernquist,
  Lancaster et~al.}}]{Mocz_2023}
\bibinfo{author}{\bibfnamefont{P.}~\bibnamefont{Mocz}},
  \bibinfo{author}{\bibfnamefont{A.}~\bibnamefont{Fialkov}},
  \bibinfo{author}{\bibfnamefont{M.}~\bibnamefont{Vogelsberger}},
  \bibinfo{author}{\bibfnamefont{M.}~\bibnamefont{Boylan-Kolchin}},
  \bibinfo{author}{\bibfnamefont{P.-H.} \bibnamefont{Chavanis}},
  \bibinfo{author}{\bibfnamefont{M.~A.} \bibnamefont{Amin}},
  \bibinfo{author}{\bibfnamefont{S.}~\bibnamefont{Bose}},
  \bibinfo{author}{\bibfnamefont{T.}~\bibnamefont{Dome}},
  \bibinfo{author}{\bibfnamefont{L.}~\bibnamefont{Hernquist}},
  \bibinfo{author}{\bibfnamefont{L.}~\bibnamefont{Lancaster}},
  \bibnamefont{et~al.}, \bibinfo{journal}{Monthly Notices of the Royal
  Astronomical Society} \textbf{\bibinfo{volume}{521}}, \bibinfo{pages}{2608}
  (\bibinfo{year}{2023}), ISSN \bibinfo{issn}{0035-8711},
  \urlprefix\url{https://academic.oup.com/mnras/article/521/2/2608/7070733}.

\bibitem[{\citenamefont{Chakrabarti et~al.}(2022)\citenamefont{Chakrabarti,
  Dave, Dutta, and Goswami}}]{Chakrabarti2022}
\bibinfo{author}{\bibfnamefont{S.}~\bibnamefont{Chakrabarti}},
  \bibinfo{author}{\bibfnamefont{B.}~\bibnamefont{Dave}},
  \bibinfo{author}{\bibfnamefont{K.}~\bibnamefont{Dutta}}, \bibnamefont{and}
  \bibinfo{author}{\bibfnamefont{G.}~\bibnamefont{Goswami}},
  \bibinfo{journal}{Journal of Cosmology and Astroparticle Physics}
  \textbf{\bibinfo{volume}{2022}}, \bibinfo{pages}{074} (\bibinfo{year}{2022}),
  \urlprefix\url{https://dx.doi.org/10.1088/1475-7516/2022/09/074}.

\bibitem[{\citenamefont{Dave and Goswami}(2023{\natexlab{a}})}]{Dave2023}
\bibinfo{author}{\bibfnamefont{B.}~\bibnamefont{Dave}} \bibnamefont{and}
  \bibinfo{author}{\bibfnamefont{G.}~\bibnamefont{Goswami}},
  \bibinfo{journal}{Journal of Cosmology and Astroparticle Physics}
  \textbf{\bibinfo{volume}{2023}}, \bibinfo{pages}{015}
  (\bibinfo{year}{2023}{\natexlab{a}}),
  \urlprefix\url{https://dx.doi.org/10.1088/1475-7516/2023/07/015}.

\bibitem[{\citenamefont{Dave and Goswami}(2023{\natexlab{b}})}]{Dave:2023egr}
\bibinfo{author}{\bibfnamefont{B.}~\bibnamefont{Dave}} \bibnamefont{and}
  \bibinfo{author}{\bibfnamefont{G.}~\bibnamefont{Goswami}}
  (\bibinfo{year}{2023}{\natexlab{b}}), \eprint{2310.19664}.

\bibitem[{\citenamefont{Chavanis}(2012)}]{Chavanis:2011uv}
\bibinfo{author}{\bibfnamefont{P.-H.} \bibnamefont{Chavanis}},
  \bibinfo{journal}{Astron. Astrophys.} \textbf{\bibinfo{volume}{537}},
  \bibinfo{pages}{A127} (\bibinfo{year}{2012}), \eprint{1103.2698}.

\bibitem[{\citenamefont{Brax et~al.}(2020)\citenamefont{Brax, Cembranos, and
  Valageas}}]{Brax:2020oye}
\bibinfo{author}{\bibfnamefont{P.}~\bibnamefont{Brax}},
  \bibinfo{author}{\bibfnamefont{J.~A.~R.} \bibnamefont{Cembranos}},
  \bibnamefont{and} \bibinfo{author}{\bibfnamefont{P.}~\bibnamefont{Valageas}},
  \bibinfo{journal}{Phys. Rev. D} \textbf{\bibinfo{volume}{102}},
  \bibinfo{pages}{083012} (\bibinfo{year}{2020}), \eprint{2007.04638}.

\bibitem[{\citenamefont{{Danzmann} and {LISA Study Team}}(1996)}]{Danzman-1996}
\bibinfo{author}{\bibfnamefont{K.}~\bibnamefont{{Danzmann}}} \bibnamefont{and}
  \bibinfo{author}{\bibnamefont{{LISA Study Team}}},
  \bibinfo{journal}{Classical and Quantum Gravity}
  \textbf{\bibinfo{volume}{13}}, \bibinfo{pages}{A247} (\bibinfo{year}{1996}).

\bibitem[{\citenamefont{Amaro-Seoane et~al.}(2017)}]{LISA:2017pwj}
\bibinfo{author}{\bibfnamefont{P.}~\bibnamefont{Amaro-Seoane}}
  \bibnamefont{et~al.} (\bibinfo{collaboration}{LISA}) (\bibinfo{year}{2017}),
  \eprint{1702.00786}.

\bibitem[{\citenamefont{Seoane et~al.}(2023)}]{LISA:2022yao}
\bibinfo{author}{\bibfnamefont{P.~A.} \bibnamefont{Seoane}}
  \bibnamefont{et~al.} (\bibinfo{collaboration}{LISA}),
  \bibinfo{journal}{Living Rev. Rel.} \textbf{\bibinfo{volume}{26}},
  \bibinfo{pages}{2} (\bibinfo{year}{2023}), \eprint{2203.06016}.

\bibitem[{\citenamefont{{Seto} et~al.}(2001)\citenamefont{{Seto}, {Kawamura},
  and {Nakamura}}}]{Seto-2001}
\bibinfo{author}{\bibfnamefont{N.}~\bibnamefont{{Seto}}},
  \bibinfo{author}{\bibfnamefont{S.}~\bibnamefont{{Kawamura}}},
  \bibnamefont{and}
  \bibinfo{author}{\bibfnamefont{T.}~\bibnamefont{{Nakamura}}},
  \bibinfo{journal}{\prl} \textbf{\bibinfo{volume}{87}}, \bibinfo{eid}{221103}
  (\bibinfo{year}{2001}), \eprint{astro-ph/0108011}.

\bibitem[{\citenamefont{{Nakamura} et~al.}(2016)\citenamefont{{Nakamura},
  {Ando}, {Kinugawa}, {Nakano}, {Eda}, {Sato}, {Musha}, {Akutsu}, {Tanaka},
  {Seto} et~al.}}]{Nakamura-2016}
\bibinfo{author}{\bibfnamefont{T.}~\bibnamefont{{Nakamura}}},
  \bibinfo{author}{\bibfnamefont{M.}~\bibnamefont{{Ando}}},
  \bibinfo{author}{\bibfnamefont{T.}~\bibnamefont{{Kinugawa}}},
  \bibinfo{author}{\bibfnamefont{H.}~\bibnamefont{{Nakano}}},
  \bibinfo{author}{\bibfnamefont{K.}~\bibnamefont{{Eda}}},
  \bibinfo{author}{\bibfnamefont{S.}~\bibnamefont{{Sato}}},
  \bibinfo{author}{\bibfnamefont{M.}~\bibnamefont{{Musha}}},
  \bibinfo{author}{\bibfnamefont{T.}~\bibnamefont{{Akutsu}}},
  \bibinfo{author}{\bibfnamefont{T.}~\bibnamefont{{Tanaka}}},
  \bibinfo{author}{\bibfnamefont{N.}~\bibnamefont{{Seto}}},
  \bibnamefont{et~al.}, \bibinfo{journal}{Progress of Theoretical and
  Experimental Physics} \textbf{\bibinfo{volume}{2016}}, \bibinfo{eid}{093E01}
  (\bibinfo{year}{2016}), \eprint{1607.00897}.

\bibitem[{\citenamefont{Kawamura et~al.}(2021)}]{Kawamura:2020pcg}
\bibinfo{author}{\bibfnamefont{S.}~\bibnamefont{Kawamura}}
  \bibnamefont{et~al.}, \bibinfo{journal}{PTEP}
  \textbf{\bibinfo{volume}{2021}}, \bibinfo{pages}{05A105}
  (\bibinfo{year}{2021}), \eprint{2006.13545}.

\bibitem[{\citenamefont{Vicente and Cardoso}(2022)}]{Vicente:2022ivh}
\bibinfo{author}{\bibfnamefont{R.}~\bibnamefont{Vicente}} \bibnamefont{and}
  \bibinfo{author}{\bibfnamefont{V.}~\bibnamefont{Cardoso}},
  \bibinfo{journal}{Phys. Rev. D} \textbf{\bibinfo{volume}{105}},
  \bibinfo{pages}{083008} (\bibinfo{year}{2022}), \eprint{2201.08854}.

\bibitem[{\citenamefont{Cardoso and Maselli}(2020)}]{Cardoso:2019rou}
\bibinfo{author}{\bibfnamefont{V.}~\bibnamefont{Cardoso}} \bibnamefont{and}
  \bibinfo{author}{\bibfnamefont{A.}~\bibnamefont{Maselli}},
  \bibinfo{journal}{Astron. Astrophys.} \textbf{\bibinfo{volume}{644}},
  \bibinfo{pages}{A147} (\bibinfo{year}{2020}), \eprint{1909.05870}.

\bibitem[{\citenamefont{Bamber et~al.}(2022)\citenamefont{Bamber, Aurrekoetxea,
  Clough, and Ferreira}}]{Bamber:2022aa}
\bibinfo{author}{\bibfnamefont{J.}~\bibnamefont{Bamber}},
  \bibinfo{author}{\bibfnamefont{J.~C.} \bibnamefont{Aurrekoetxea}},
  \bibinfo{author}{\bibfnamefont{K.}~\bibnamefont{Clough}}, \bibnamefont{and}
  \bibinfo{author}{\bibfnamefont{P.~G.} \bibnamefont{Ferreira}}
  (\bibinfo{year}{2022}), \eprint{2210.09254},
  \urlprefix\url{https://arxiv.org/pdf/2210.09254.pdf}.

\bibitem[{\citenamefont{Traykova et~al.}(2023)\citenamefont{Traykova, Vicente,
  Clough, Helfer, Berti, Ferreira, and Hui}}]{Traykova:2023aa}
\bibinfo{author}{\bibfnamefont{D.}~\bibnamefont{Traykova}},
  \bibinfo{author}{\bibfnamefont{R.}~\bibnamefont{Vicente}},
  \bibinfo{author}{\bibfnamefont{K.}~\bibnamefont{Clough}},
  \bibinfo{author}{\bibfnamefont{T.}~\bibnamefont{Helfer}},
  \bibinfo{author}{\bibfnamefont{E.}~\bibnamefont{Berti}},
  \bibinfo{author}{\bibfnamefont{P.~G.} \bibnamefont{Ferreira}},
  \bibnamefont{and} \bibinfo{author}{\bibfnamefont{L.}~\bibnamefont{Hui}}
  (\bibinfo{year}{2023}), \eprint{2305.10492},
  \urlprefix\url{https://arxiv.org/pdf/2305.10492.pdf}.

\bibitem[{\citenamefont{Boudon et~al.}(2023)\citenamefont{Boudon, Brax,
  Valageas, and Wong}}]{Boudon:2023vzl}
\bibinfo{author}{\bibfnamefont{A.}~\bibnamefont{Boudon}},
  \bibinfo{author}{\bibfnamefont{P.}~\bibnamefont{Brax}},
  \bibinfo{author}{\bibfnamefont{P.}~\bibnamefont{Valageas}}, \bibnamefont{and}
  \bibinfo{author}{\bibfnamefont{L.~K.} \bibnamefont{Wong}}
  (\bibinfo{year}{2023}), \eprint{2305.18540}.

\bibitem[{\citenamefont{Aurrekoetxea et~al.}(2023)\citenamefont{Aurrekoetxea,
  Clough, Bamber, and Ferreira}}]{Aurrekoetxea:2023aa}
\bibinfo{author}{\bibfnamefont{J.~C.} \bibnamefont{Aurrekoetxea}},
  \bibinfo{author}{\bibfnamefont{K.}~\bibnamefont{Clough}},
  \bibinfo{author}{\bibfnamefont{J.}~\bibnamefont{Bamber}}, \bibnamefont{and}
  \bibinfo{author}{\bibfnamefont{P.~G.} \bibnamefont{Ferreira}}
  (\bibinfo{year}{2023}), \eprint{2311.18156},
  \urlprefix\url{https://arxiv.org/pdf/2311.18156.pdf}.

\bibitem[{\citenamefont{Khmelnitsky and Rubakov}(2014)}]{Khmelnitsky:2013lxt}
\bibinfo{author}{\bibfnamefont{A.}~\bibnamefont{Khmelnitsky}} \bibnamefont{and}
  \bibinfo{author}{\bibfnamefont{V.}~\bibnamefont{Rubakov}},
  \bibinfo{journal}{JCAP} \textbf{\bibinfo{volume}{02}}, \bibinfo{pages}{019}
  (\bibinfo{year}{2014}), \eprint{1309.5888}.

\bibitem[{\citenamefont{Chandrasekhar}(1943)}]{Chandrasekhar:1943ys}
\bibinfo{author}{\bibfnamefont{S.}~\bibnamefont{Chandrasekhar}},
  \bibinfo{journal}{Astrophys. J.} \textbf{\bibinfo{volume}{97}},
  \bibinfo{pages}{255} (\bibinfo{year}{1943}).

\bibitem[{\citenamefont{Barausse et~al.}(2014)\citenamefont{Barausse, Cardoso,
  and Pani}}]{Barausse:2014tra}
\bibinfo{author}{\bibfnamefont{E.}~\bibnamefont{Barausse}},
  \bibinfo{author}{\bibfnamefont{V.}~\bibnamefont{Cardoso}}, \bibnamefont{and}
  \bibinfo{author}{\bibfnamefont{P.}~\bibnamefont{Pani}},
  \bibinfo{journal}{Phys. Rev. D} \textbf{\bibinfo{volume}{89}},
  \bibinfo{pages}{104059} (\bibinfo{year}{2014}), \eprint{1404.7149}.

\bibitem[{\citenamefont{Lamberts et~al.}(2019)\citenamefont{Lamberts, Blunt,
  Littenberg, Garrison-Kimmel, Kupfer, and Sanderson}}]{Lamberts:2019nyk}
\bibinfo{author}{\bibfnamefont{A.}~\bibnamefont{Lamberts}},
  \bibinfo{author}{\bibfnamefont{S.}~\bibnamefont{Blunt}},
  \bibinfo{author}{\bibfnamefont{T.~B.} \bibnamefont{Littenberg}},
  \bibinfo{author}{\bibfnamefont{S.}~\bibnamefont{Garrison-Kimmel}},
  \bibinfo{author}{\bibfnamefont{T.}~\bibnamefont{Kupfer}}, \bibnamefont{and}
  \bibinfo{author}{\bibfnamefont{R.~E.} \bibnamefont{Sanderson}},
  \bibinfo{journal}{Mon. Not. Roy. Astron. Soc.}
  \textbf{\bibinfo{volume}{490}}, \bibinfo{pages}{5888} (\bibinfo{year}{2019}),
  \eprint{1907.00014}.

\bibitem[{\citenamefont{Seto}(2022)}]{Seto:2022iuf}
\bibinfo{author}{\bibfnamefont{N.}~\bibnamefont{Seto}}, \bibinfo{journal}{Phys.
  Rev. Lett.} \textbf{\bibinfo{volume}{128}}, \bibinfo{pages}{041101}
  (\bibinfo{year}{2022}), \eprint{2201.03685}.

\bibitem[{\citenamefont{Poisson and Will}(2014)}]{poisson_will_2014}
\bibinfo{author}{\bibfnamefont{E.}~\bibnamefont{Poisson}} \bibnamefont{and}
  \bibinfo{author}{\bibfnamefont{C.~M.} \bibnamefont{Will}},
  \emph{\bibinfo{title}{Gravity: Newtonian, Post-Newtonian, Relativistic}}
  (\bibinfo{publisher}{Cambridge University Press}, \bibinfo{year}{2014}).

\bibitem[{\citenamefont{Poisson and Will}(1995)}]{Poisson:1995ef}
\bibinfo{author}{\bibfnamefont{E.}~\bibnamefont{Poisson}} \bibnamefont{and}
  \bibinfo{author}{\bibfnamefont{C.~M.} \bibnamefont{Will}},
  \bibinfo{journal}{Phys. Rev. D} \textbf{\bibinfo{volume}{52}},
  \bibinfo{pages}{848} (\bibinfo{year}{1995}), \eprint{gr-qc/9502040}.

\bibitem[{\citenamefont{Vallisneri}(2008)}]{Vallisneri:2007ev}
\bibinfo{author}{\bibfnamefont{M.}~\bibnamefont{Vallisneri}},
  \bibinfo{journal}{Phys. Rev. D} \textbf{\bibinfo{volume}{77}},
  \bibinfo{pages}{042001} (\bibinfo{year}{2008}), \eprint{gr-qc/0703086}.

\bibitem[{\citenamefont{Klein et~al.}(2016)}]{Klein:2015hvg}
\bibinfo{author}{\bibfnamefont{A.}~\bibnamefont{Klein}} \bibnamefont{et~al.},
  \bibinfo{journal}{Phys. Rev. D} \textbf{\bibinfo{volume}{93}},
  \bibinfo{pages}{024003} (\bibinfo{year}{2016}), \eprint{1511.05581}.

\bibitem[{\citenamefont{Bonetti et~al.}(2019)\citenamefont{Bonetti, Sesana,
  Haardt, Barausse, and Colpi}}]{Bonetti:2018tpf}
\bibinfo{author}{\bibfnamefont{M.}~\bibnamefont{Bonetti}},
  \bibinfo{author}{\bibfnamefont{A.}~\bibnamefont{Sesana}},
  \bibinfo{author}{\bibfnamefont{F.}~\bibnamefont{Haardt}},
  \bibinfo{author}{\bibfnamefont{E.}~\bibnamefont{Barausse}}, \bibnamefont{and}
  \bibinfo{author}{\bibfnamefont{M.}~\bibnamefont{Colpi}},
  \bibinfo{journal}{Mon. Not. Roy. Astron. Soc.}
  \textbf{\bibinfo{volume}{486}}, \bibinfo{pages}{4044} (\bibinfo{year}{2019}),
  \eprint{1812.01011}.

\bibitem[{\citenamefont{Dayal et~al.}(2019)\citenamefont{Dayal, Rossi,
  Shiralilou, Piana, Choudhury, and Volonteri}}]{Dayal:2018gwg}
\bibinfo{author}{\bibfnamefont{P.}~\bibnamefont{Dayal}},
  \bibinfo{author}{\bibfnamefont{E.~M.} \bibnamefont{Rossi}},
  \bibinfo{author}{\bibfnamefont{B.}~\bibnamefont{Shiralilou}},
  \bibinfo{author}{\bibfnamefont{O.}~\bibnamefont{Piana}},
  \bibinfo{author}{\bibfnamefont{T.~R.} \bibnamefont{Choudhury}},
  \bibnamefont{and}
  \bibinfo{author}{\bibfnamefont{M.}~\bibnamefont{Volonteri}},
  \bibinfo{journal}{Mon. Not. Roy. Astron. Soc.}
  \textbf{\bibinfo{volume}{486}}, \bibinfo{pages}{2336} (\bibinfo{year}{2019}),
  \eprint{1810.11033}.

\bibitem[{\citenamefont{Ricarte and Natarajan}(2018)}]{Ricarte:2018}
\bibinfo{author}{\bibfnamefont{A.}~\bibnamefont{Ricarte}} \bibnamefont{and}
  \bibinfo{author}{\bibfnamefont{P.}~\bibnamefont{Natarajan}},
  \bibinfo{journal}{Monthly Notices of the Royal Astronomical Society}
  \textbf{\bibinfo{volume}{481}}, \bibinfo{pages}{3278} (\bibinfo{year}{2018}),
  ISSN \bibinfo{issn}{0035-8711},
  \eprint{https://academic.oup.com/mnras/article-pdf/481/3/3278/25834715/sty2448.pdf},
  \urlprefix\url{https://doi.org/10.1093/mnras/sty2448}.

\bibitem[{\citenamefont{Ruiter et~al.}(2010)\citenamefont{Ruiter, Belczynski,
  Benacquista, Larson, and Williams}}]{Ruiter:2007xx}
\bibinfo{author}{\bibfnamefont{A.~J.} \bibnamefont{Ruiter}},
  \bibinfo{author}{\bibfnamefont{K.}~\bibnamefont{Belczynski}},
  \bibinfo{author}{\bibfnamefont{M.}~\bibnamefont{Benacquista}},
  \bibinfo{author}{\bibfnamefont{S.~L.} \bibnamefont{Larson}},
  \bibnamefont{and} \bibinfo{author}{\bibfnamefont{G.}~\bibnamefont{Williams}},
  \bibinfo{journal}{Astrophys. J.} \textbf{\bibinfo{volume}{717}},
  \bibinfo{pages}{1006} (\bibinfo{year}{2010}), \eprint{0705.3272}.

\bibitem[{\citenamefont{Breivik et~al.}(2020)}]{Breivik:2019lmt}
\bibinfo{author}{\bibfnamefont{K.}~\bibnamefont{Breivik}} \bibnamefont{et~al.},
  \bibinfo{journal}{Astrophys. J.} \textbf{\bibinfo{volume}{898}},
  \bibinfo{pages}{71} (\bibinfo{year}{2020}), \eprint{1911.00903}.

\bibitem[{\citenamefont{Abbott et~al.}(2019)}]{LIGOScientific:2018mvr}
\bibinfo{author}{\bibfnamefont{B.~P.} \bibnamefont{Abbott}}
  \bibnamefont{et~al.} (\bibinfo{collaboration}{LIGO Scientific, Virgo}),
  \bibinfo{journal}{Phys. Rev. X} \textbf{\bibinfo{volume}{9}},
  \bibinfo{pages}{031040} (\bibinfo{year}{2019}), \eprint{1811.12907}.

\bibitem[{\citenamefont{{Kinugawa} et~al.}(2022)\citenamefont{{Kinugawa},
  {Takeda}, {Tanikawa}, and {Yamaguchi}}}]{Kinugawa2022}
\bibinfo{author}{\bibfnamefont{T.}~\bibnamefont{{Kinugawa}}},
  \bibinfo{author}{\bibfnamefont{H.}~\bibnamefont{{Takeda}}},
  \bibinfo{author}{\bibfnamefont{A.}~\bibnamefont{{Tanikawa}}},
  \bibnamefont{and}
  \bibinfo{author}{\bibfnamefont{H.}~\bibnamefont{{Yamaguchi}}},
  \bibinfo{journal}{\apj} \textbf{\bibinfo{volume}{938}}, \bibinfo{eid}{52}
  (\bibinfo{year}{2022}), \eprint{1910.01063}.

\end{thebibliography}

\end{document}